\makeatletter
\newcommand{\dontusepackage}[2][]{%
  \@namedef{ver@#2.sty}{9999/12/31}%
  \@namedef{opt@#2.sty}{#1}}
\makeatother
\dontusepackage{subfigure}

\documentclass[]{article}

\usepackage{lmodern}
\usepackage{amssymb,amsmath}
\usepackage{ifxetex,ifluatex}
\usepackage[usenames,dvipsnames]{color}
\usepackage{fixltx2e} 
\ifnum 0\ifxetex 1\fi\ifluatex 1\fi=0 
  \usepackage[T1]{fontenc}
  \usepackage[utf8]{inputenc}
\else 
  \ifxetex
    \usepackage{mathspec}
    \usepackage{xltxtra,xunicode}
  \else
    \usepackage{fontspec}
  \fi
  \defaultfontfeatures{Mapping=tex-text,Scale=MatchLowercase}
  
\fi
\IfFileExists{upquote.sty}{\usepackage{upquote}}{}
\IfFileExists{microtype.sty}{%
\usepackage{microtype}
\UseMicrotypeSet[protrusion]{basicmath} 
}{}
\usepackage[margin=1.0in,bottom=1.5in]{geometry}
\usepackage[square,numbers,sort&compress]{natbib}
\usepackage{listings}
\usepackage{longtable,booktabs}
\usepackage{graphicx}
\makeatletter
\def\maxwidth{\ifdim\Gin@nat@width>\linewidth\linewidth\else\Gin@nat@width\fi}
\def\maxheight{\ifdim\Gin@nat@height>\textheight\textheight\else\Gin@nat@height\fi}
\makeatother
\setkeys{Gin}{width=\maxwidth,height=\maxheight,keepaspectratio}
\usepackage{caption}
\usepackage{float}
\usepackage{subfig}
\usepackage{algorithm} 
\let\scholmdAlgorithm\algorithm
\let\endscholmdAlgorithm\endalgorithm
\let\algorithm\relax \let\endalgorithm\relax

{
 \catcode`\*=11\relax
 \global\let\scholmdAlgorithm*\algorithm*
 \global\let\endscholmdAlgorithm*\endalgorithm*
 \global\let\algorithm*\relax 
 \global\let\endalgorithm*\relax
}
\ifxetex
  \usepackage[setpagesize=false, 
              unicode=false, 
              xetex]{hyperref}
\else
  \usepackage[unicode=true]{hyperref}
\fi
\hypersetup{breaklinks=true,
            bookmarks=true,
            pdfauthor={},
            pdftitle={Neural network augmented wave-equation simulation},
            colorlinks=true,
            citecolor=black,
            urlcolor=blue,
            linkcolor=black,
            pdfborder={0 0 0}}
\usepackage[preprint]{neurips_2019}
\usepackage{url}
\usepackage{booktabs}
\usepackage{amsfonts}
\usepackage{nicefrac}
\usepackage{microtype}

\title{Neural network augmented wave-equation simulation}
\author{Ali Siahkoohi, Mathias Louboutin, and Felix J. Herrmann\\School of
Computational Science and Engineering\\Georgia Institute of
Technology\\\texttt{\{alisk,\phantom{\ }mlouboutin3,\phantom{\ }felix.herrmann\}@gatech.edu}}
\date{}

\begin{document}
\maketitle
\begin{abstract}
Accurate forward modeling is important for solving inverse problems. An
inaccurate wave-equation simulation, as a forward operator, will offset
the results obtained via inversion. In this work, we consider the case
where we deal with incomplete physics. One proxy of incomplete physics
is an inaccurate discretization of Laplacian in simulation of wave
equation via finite-difference method. We exploit intrinsic one-to-one
similarities between timestepping algorithm with Convolutional Neural
Networks (CNNs), and propose to intersperse CNNs between low-fidelity
timesteps. Augmenting neural networks with low-fidelity timestepping
algorithms may allow us to take large timesteps while limiting the
numerical dispersion artifacts. While simulating the wave-equation with
low-fidelity timestepping algorithm, by correcting the wavefield several
time during propagation, we hope to limit the numerical dispersion
artifact introduced by a poor discretization of the Laplacian. As a
proof of concept, we demonstrate this principle by correcting for
numerical dispersion by keeping the velocity model fixed, and varying
the source locations to generate training and testing pairs for our
supervised learning algorithm.
\end{abstract}

\section{Introduction}\label{introduction}

In inverse problem, we heavily rely on having an accurate forward
modeling operators. Often, we can not afford being physically or
numerically accurate. In other words, being numerically inaccurate can
be due to computationally complexity of accurate methods, or incomplete
knowledge of the underlying data generation process. In either case,
motivated by \citet{ruthotto2018deep}, we propose to intersperse CNNs
between timestepping algorithm for simulating acoustic wave-equation. We
mimic incomplete/inaccurate physics by simulating wave equation with
finite-difference method, while utilizing a poor (second-order)
discretization of Laplacian.

Conventional method for solving partial differential equations (PDEs),
e.g., finite-difference and finite-element method, given enough
computational recourses, are able to simulate high-fidelity solutions to
PDEs. On one hand, as long as the Courant--Friedrichs--Lewy conditions
for stability are satisfied, finite-difference methods are able to
compute solutions to PDE, regardless of medium parameters, with
arbitrary precision. On the other hand, finite-element method requires
careful meshing of the medium in order to carry out the simulation.
Another motivation behind this work is to exploit the fact that in
seismic, wave simulations are usually carried out for specific families
of velocity models and source/receiver distributions. We hope our
proposed method meets halfway between two mentioned extremes---i.e.,
being too generic (finite-difference method) and being too problem
specific (finite-element method).

There are several attempts to exploit learning methods in wave-equation
simulation. \citet{raissi2018deep} approximates the solution to a
nonlinear PDE with a neural network. The neural network, given points on
the computational grid as input, computes the solution of PDE. Training
data is obtained by computing the solution of PDE on several points.
\citet{moseley2018fast} completely ignore the Laplacian and they solely
rely on predicting the next timestep from the previous two timesteps by
learning the action of the spatially varying velocity and Laplacian.
While possible in principle, their approach needs to train for long
times to provide reasonable simulations on relatively simple models.
\citet{siahkoohi2019transfer} instead of ignoring the physics, relies on
low-fidelity wave-equation simulation, and by exploiting transfer
learning, they utilize a single CNN to correct wavefield snapshots
simulated on a ``nearby'' velocity mode for numerical dispersion at any
given timestep. In this work, we extend ideas in
\citet{siahkoohi2019transfer} and propose using multiple CNNs,
interspersed between low-fidelity timesteps. Finally,
\citet{rizzuti2019EAGElis} propose interspersing Krylov-subspace
iterations and neural nets while inverting the Helmholtz equation. They
show improvement in convergence by ``propagating'' an approximated
wavefield, obtained from a limited number of iterations, with the aid of
a trained convolutional neural net. This technique can be seen as the
frequency-domain counterpart of our proposed method.

Our paper is organized as follows. First, we describe our approach in
detail by first, describing our formulation for learned wave simulation.
Next, we introduce our training objective function. Due to dependencies
of CNN parameters, we devised an training heuristic that we describe.
Before explaining our numerical experiments, we state used CNN
architecture and training details. Next, we describe our three numerical
experiments we conduct and discuss effectiveness of the proposed method.

\section{Theory}\label{theory}

We describe how we augment low-fidelity physics with learning techniques
to handle incomplete and/or inaccurate physics, where the low-fidelity
physics is modeled via finite-difference method with a poor
discretization of the Laplacian. To ensure accuracy, the temporal and
spatial discretization in high-fidelity wave-equation simulations have
to be chosen very fine, typically one to two orders of magnitude smaller
than Nyquist sampling rate. As mentioned earlier, we will utilize a poor
discretization (only second order) for the Laplacian to carry out
low-fidelity wave-equation simulations, but the scheme can be extended
to other proxies of incomplete or inaccurate physics.

\subsection{Simulations by
timestepping}\label{simulations-by-timestepping}

After discretization of the acoustic wave equation, a single timestep of
of scalar wavefields, simulated on $0 \leq t \leq T$, can be written as
below:
\begin{equation}
    \mathbf{u}_{j+1} = 2\mathbf{u}_{j}-\mathbf{u}_{j-1}+\delta t^2\mathbf{c}^2 \mathbf{\Delta}\mathbf{u}_j, \quad j=0,1,  \ldots , N-1 ,
\label{high-fidelity}
\end{equation}
 where $\mathbf{u}_{j}$ is the high-fidelity scalar wavefield at
$j^{\text{th}}$ timestep, $\delta t$ is the temporal discretization
(timestep), $\mathbf{c}$ is the spatially varying velocity in the
medium, and $\mathbf{\Delta}$ is the high-order discretization of
Laplacian. Similar to Equation~\ref{high-fidelity}, the low-fidelity
timestepping equation can be formulated as
\begin{equation}
\mathbf{\bar{u}}_{j+1} = 2\mathbf{\bar{u}}_{j}-\mathbf{\bar{u}}_{j-1}+\delta T^2\mathbf{\bar{c}}^2 \mathbf{\bar{\Delta}}\mathbf{\bar{u}}_j,\quad j=0, 1, \ldots , M-1,
\label{low-fidelity}
\end{equation}
 where $\mathbf{\bar{u}}_{j}$ is the low-fidelity scalar wavefield,
$\delta T$ is the coarse timestep, $\mathbf{\bar{c}}$ is the coarse
spatially varying velocity, and $\mathbf{\bar{\Delta}}$ is the coarse
(only second second order) discretized Laplacian.

\begin{figure}
\centering
\subfloat[\label{hf-block}]{\includegraphics[width=0.500\hsize]{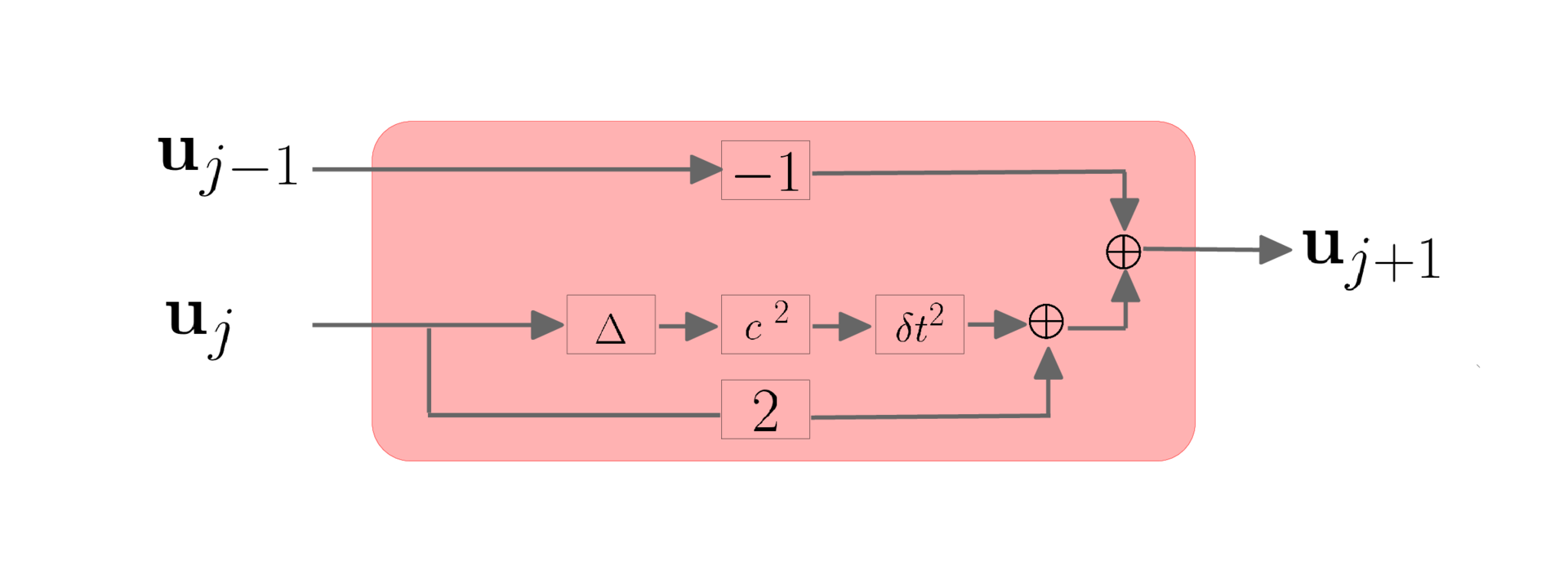}}
\subfloat[\label{lf-block}]{\includegraphics[width=0.500\hsize]{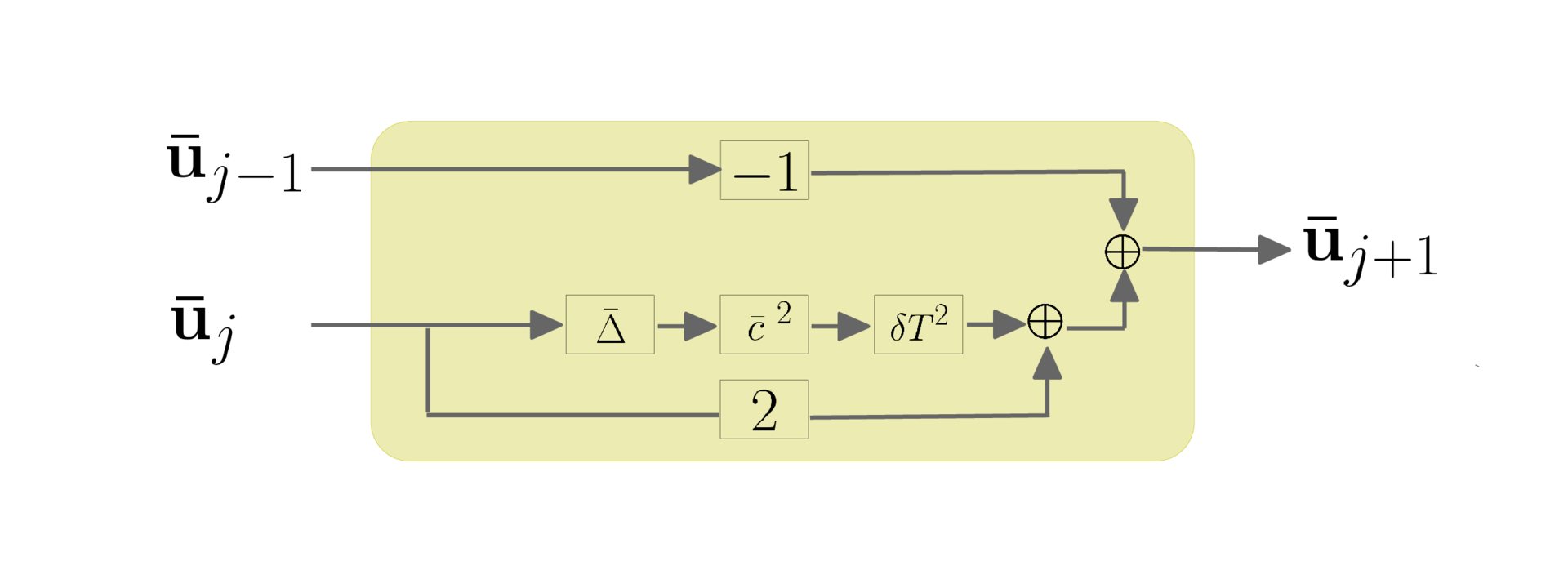}}
\caption{Comparing a single low and high-fidelity timestep. a)
High-fidelity timestep. b) Low-fidelity timestep.}\label{Low-High}
\end{figure}

\begin{figure}
\centering
\subfloat[\label{high-fidelity-step}]{\includegraphics[width=1.000\hsize]{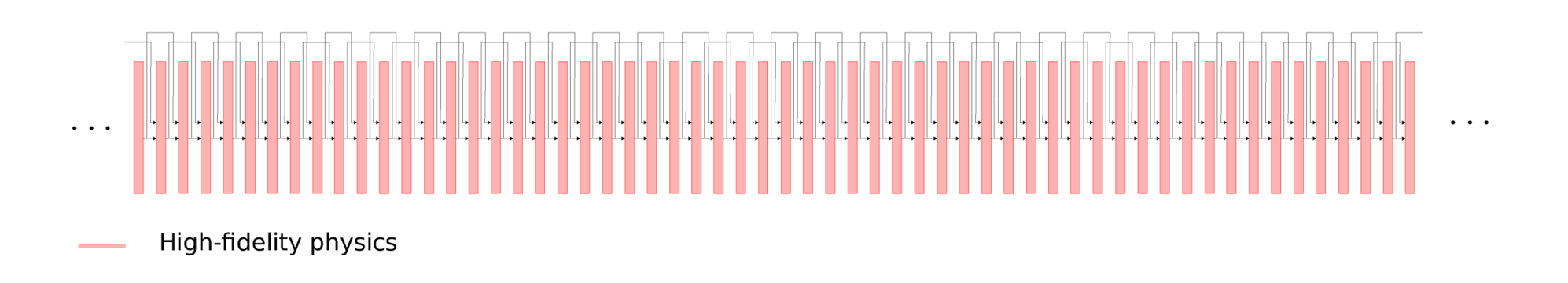}}
\\
\subfloat[\label{low-fidelity-step}]{\includegraphics[width=0.909\hsize]{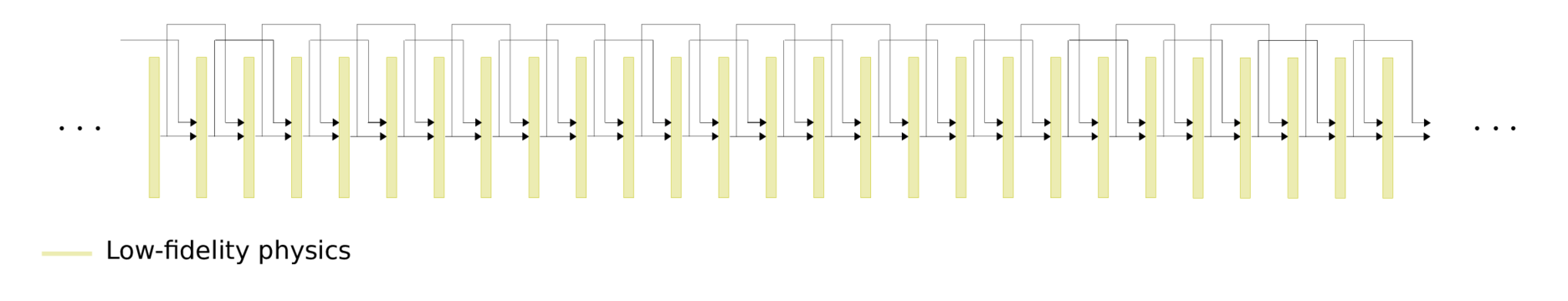}}
\caption{Comparing low and high-fidelity discretized wave equation
simulations. a) High-fidelity simulation. b) Low-fidelity
simulation.}\label{Stepping}
\end{figure}

Motivated by \citet{ruthotto2018deep}, we consider every timestep as a
single layer in a neural network, where the discretized Laplacian is a
linear operator, followed by the (nonlinear) action of the spatial
varying velocity. Moreover, noticing the additional terms in the
Equation~\ref{high-fidelity}, each timestep is similar to a residual
block introduced by \citet{szegedy2017inception}. Figures~\ref{hf-block}
and~\ref{lf-block} schematically indicate each timestep as a block,
corresponding to high- and low-fidelity discretization of wave equation.
respectively. The similarity of high- and low -fidelity timestepping
method and CNNs can be perceived from Figures~\ref{high-fidelity-step}
and low-fidelity-step, respectively, where red and yellow blocks
correspond to high- and low-fidelity timestepping equations,
respectively.

As it can be seen, high-fidelity simulation of the wave equation up to
time $t=T$ requires a lot of high-fidelity timesteps. On the other hand,
Figure~\ref{low-fidelity-step} shows that the low-fidelity simulations
can be done with much less low-fidelity timesteps, due to course time
sampling, which each timestep is cheaper than the high-fidelity
timesteps due to the coarse discretization of Laplacian. Although
computationally cheap, the low-fidelity wave-equation simulations suffer
from numerical dispersion artifacts.

\subsection{Learned wave simulations}\label{learned-wave-simulations}

\begin{figure}
\centering
\includegraphics[width=1.000\hsize]{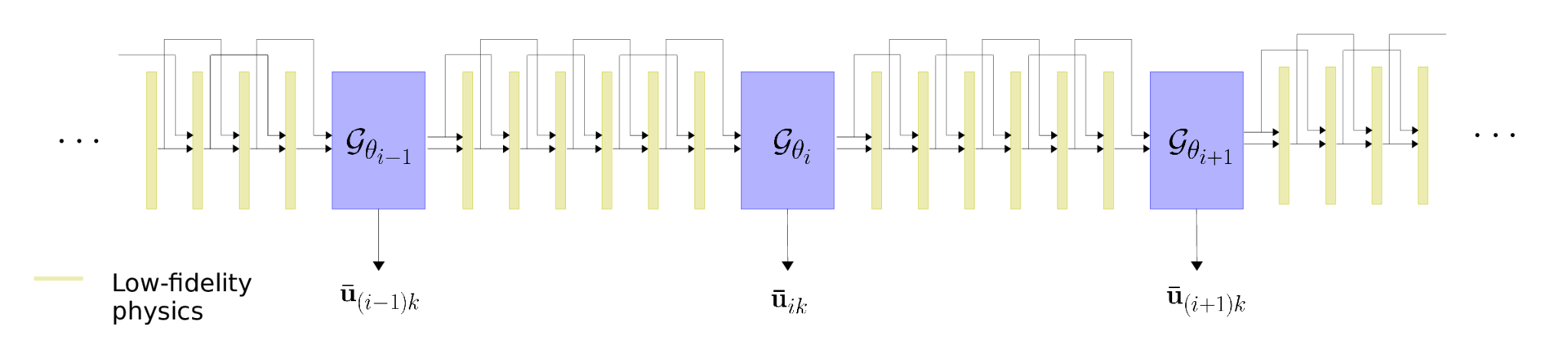}
\caption{A schematic representation of the proposed
method.}\label{proposed}
\end{figure}

Depending on the domain of application, we can assume wave simulations
are typically carried out for specific families of velocity models and
source/receiver distributions. This motivates us to deploy a data-driven
wave simulation algorithm which is coupled with low-fidelity and cheap
physics and hope to recover high-fidelity wave simulations on a family
of velocity models. In our method, we propagate the coarse-grained
wavefields according to Equation~\ref{low-fidelity} with a coarsened
Laplacian. After $k$ timesteps, where $k$ is a hyperparameter, we apply
a correction with a CNN, $\mathcal{G}_{\theta_i}$, parameterized by
$\theta_i$, to the obtained wavefield at $j^{\text{th}}$ timestep and
proceed with the timestepping. The proposed data-driven timestepping
wave simulation method is formalized in Equation~\ref{learned}.
\begin{equation}
\mathbf{\bar{u}}_{j+1} = \begin{cases}
\mathcal{G}_{\theta_i}\left( 2\mathbf{\bar{u}}_{j}-\mathbf{\bar{u}}_{j-1}+\delta T^2 \mathbf{\bar{c}}^2 \mathbf{\bar{\Delta}}\mathbf{\bar{u}}_j\right), \ i = \lfloor \frac{j}{k} \rfloor & \text{ if } j\equiv k-1 \ (\bmod \ k),   \\ 
2\mathbf{\bar{u}}_{j}-\mathbf{\bar{u}}_{j-1}+\delta T^2 \mathbf{\bar{c}}^2 \mathbf{\bar{\Delta}}\mathbf{\bar{u}}_j & \text{ else }
\end{cases}
\label{learned}
\end{equation}
 where $j=0,1, \ldots , M-1$. The schematic representation of
Equation~\ref{learned} is illustrated in Figure~\ref{proposed}. Yellow
blocks represent low-fidelity timesteps (see Equation~\ref{low-fidelity}
and Figure~\ref{lf-block}) and blue blocks correspond to CNNs,
$\mathcal{G}_{\theta_i}, \ i = 0, 1, \ldots , \lfloor \frac{M-1}{k} \rfloor$.

The CNNs correct for the effects of inaccurate physics---i.e., numerical
dispersion in our experiments, at every $k^{\text{th}}$ low-fidelity
timestep. In this work, the parameters, $\theta_i$, are not shared among
the CNNs. We have not explored the possibility of shared weights among
the CNNs in different stages of wave propagation.

Note that although parameters of the CNNs are not shared, they are not
independent---i.e., after $j^{\text{th}}$ timestep, CNN
$\mathcal{G}_{\theta_i}, i = \lfloor \frac{j}{k} \rfloor$, corrects for
errors in the wavefield introduced by low-fidelity timestepping and
imperfections present in the output of $(i-1)^{\text{th}}$ CNN, which
have been propagated through timestepping. Therefore, a small
perturbation in the parameters of a CNN in the initial stages of neural
network augmented timestepping causes noticeable differences in the
input of the CNNs in later stages of the wave propagation. The described
dependencies among the CNN parameters introduces difficulties in
optimizing the parameters of the CNNs. Below we describe our heuristic
for training the CNNs.

\subsection{Training objective}\label{training-objective}

During training, we train all the CNNs toward the high-fidelity solution
of wave-equation, at the corresponding timestep, obtained by solving
Equation~\ref{high-fidelity}. As it can be seen from
Equation~\ref{learned}, after $j^{\text{th}}$ timestep, CNN
$\mathcal{G}_{\theta_i}, i = \lfloor \frac{j}{k} \rfloor$, is tasked to
correct the effects of low-fidelity timestepping. During training,
$i^{\text{th}}$ CNN maps its input,
$2\mathbf{\bar{u}}_{j}-\mathbf{\bar{u}}_{j-1}+\delta T^2 \mathbf{\bar{c}}^2 \mathbf{\bar{\Delta}} \mathbf{\bar{u}}_j$,
to $\mathbf{u}_{j+1}$, result of $j^{\text{th}}$ timestep using
high-fidelity timestepping, obtained by Equation~\ref{high-fidelity}.

Define function $\bar{F}_k(.)$ as the action of $k$ low-fidelity
timesteps---i.e., $\bar{F}_k(.)$ represents $k$ consecutive low-fidelity
time stepping blocks, depicted in Figure~\ref{lf-block}. Clearly,
$\bar{F}_k$ is only a function of $k$, $\delta T$, $\mathbf{\bar{c}}$,
and $\mathbf{\bar{\Delta}}$. Using the defined notation, we can write
the input to $i^{\text{th}}$ CNN,
$\mathbf{\hat{u}}_i, \ i =0, 1, \ldots , \lfloor \frac{M-1}{k} \rfloor$,
as follows:
\begin{equation}
\begin{aligned}
    &\ \mathbf{\hat{u}}_i = \bar{F}_k (\mathcal{G}_{\theta_{i-1}}(\mathbf{\hat{u}}_{i-1} )), \quad i = 1, 2, \ldots , \lfloor \frac{M-1}{k} \rfloor, \\
    &\ \mathbf{\hat{u}}_0 = \bar{F}_k(\mathbf{q}), \\
\end{aligned}
\label{inputtoCNN}
\end{equation}
 where $\mathbf{q}$ is the source. Also let $\mathbf{u}_{\tau_i}$ denote
the wavefield obtained at $j^{\text{th}}$ timestep of high-fidelity
timestepping (Equation~\ref{high-fidelity}), where
$\tau_i = j = (k+1)i-1$.

The input-output pair of the $i^{\text{th}}$ CNN is
$(\mathbf{\hat{u}}_i, \mathbf{u}_{\tau_i})$. We can generated multiple
training pairs for CNNs by simulating
$(\mathbf{\hat{u}}_i, \mathbf{u}_{\tau_i})$ pairs, for various velocity
models and source locations. Assume we have $n$ pairs of of training
data for CNNs, namely,
$(\mathbf{\hat{u}}_i^{(p)}, \mathbf{u}_{\tau_i}^{(p)}), \ p = 0,1, \ldots , n-1$.
The objective function for optimizing $i^{\text{th}}$ CNN can be written
as follows:
\begin{equation}
\begin{split}
    \mathcal{L}_i = \frac{1}{n} \sum_{p=0}^{n-1} \left \| \mathcal{G}_{\theta_i}(\mathbf{\hat{u}}_i^{(p)}) - \mathbf{u}_{\tau_i}^{(p)} \right \|_1, \quad i = 0, 1, \ldots , \lfloor \frac{M-1}{k} \rfloor . \\
\end{split}
\label{trainCNN-i}
\end{equation}
 In the past, in a similar attempt, we used Generative Adversarial
Networks \citep[GANs,][]{Goodfellow2014} to train a CNN in order to
remove numerical dispersion from wavefield snapshots
\citep{siahkoohi2018deep, siahkoohi2019transfer}. In this work we choose
to use $\ell_1$ norm as the misfit function based on two reasons. First,
training GANs is computationally expensive since it requires training an
additional neural network that discerns between high-fidelity wavefield
snapshots and corrected ones. The computational complexity of the
proposed method in this work is significantly higher than our previous
attempts \citep{siahkoohi2018deep, siahkoohi2019transfer}, because it
involves training multiple CNNs. Based on the mentioned facts, for
limiting the computation time we chose to use $\ell_1$ misfit function.
Second, motivated by a numerical experiment performed by
\citet{hu2019render4completion}, $\ell_1$ norm misfit function yields
the second best results after the misfit function utilizing a
combination of GANs and $\ell_1$ norm misfit. In the next section, we
describe our heuristic for training the CNNs.

\subsection{Training heuristic}\label{training-heuristic}

To overcome complexities caused by dependencies between parameters of
CNNs, we optimize the objective functions $\mathcal{L}_i$ with a
heuristic described below. We minimize
$\mathcal{L}_i, \ i = 0, 1, \ldots , \lfloor \frac{M-1}{k} \rfloor$,
with respect to $\theta_i$, respectively. In other words, we minimize
$\mathcal{L}_i$ with respect to $\theta_i$, by keeping the rest of the
parameters fixed. We keep updating all the set of parameters, in a
cyclic fashion---i.e., once we updated all the parameters,
$\mathcal{L}_i, \ i = 0, 1, \ldots , \lfloor \frac{M-1}{k} \rfloor$, we
start over and update them again, in order, until a stopping criteria is
achieved. We will describe the stopping criteria used in our experiments
later.

We minimize objective functions the above objectives~\ref{trainCNN-i}
with a variant of Stochastic Gradient Descent known as the Adam
optimizer \citep{kingma2015adam} with momentum parameter $\beta = 0.9$
and a linearly decaying stepsize with initial value
$\mu = 2 \times 10^{-4}$ for both the generator and discriminator
networks. During each iteration of Adam, the gradient $\mathcal{L}_i$ is
approximated by a single randomly selected training pair. These pairs
are selected without replacement. Once all the training pairs have been
selected, we start over by randomly picking training pairs, without
replacement from the entire training set.

The optimization carries out for a predetermined number of total
\emph{iterations}, where each iterations consists of drawing a random
training pair, without replacement, and updating parameters of a CNN.
Additionally, while optimizing $\theta_i$ by keeping the rest of the
parameters fixed, before proceeding to the next set of parameters, we
carry out the optimization to update $\theta_i$ for number of
iterations, which we refer to it as \emph{mini-iterations}.
Algorithm~\ref{alg:alg1} indicates the steps for optimizing objective
functions~\ref{trainCNN-i}.

\begin{scholmdAlgorithm}
~~1.~\textbf{INPUT:}~\\\hspace*{0.333em}\hspace*{0.333em}\hspace*{0.333em}\hspace*{0.333em}\hspace*{0.333em}\hspace*{0.333em}$MaxItr \quad$~~\texttt{//\phantom{\ }total\phantom{\ }number\phantom{\ }of\phantom{\ }iterations\phantom{\ }to\phantom{\ }carry\phantom{\ }out\phantom{\ }the\phantom{\ }optimization}\\\hspace*{0.333em}\hspace*{0.333em}\hspace*{0.333em}\hspace*{0.333em}\hspace*{0.333em}\hspace*{0.333em}$MaxMiniItr \quad$~~\texttt{//\phantom{\ }mini-iterations\phantom{\ }before\phantom{\ }proceeding\phantom{\ }to\phantom{\ }the\phantom{\ }next\phantom{\ }CNN\phantom{\ }}\\\hspace*{0.333em}\hspace*{0.333em}\hspace*{0.333em}\hspace*{0.333em}\hspace*{0.333em}\hspace*{0.333em}$\bar{F}_k(.) \quad$~\texttt{//\phantom{\ }k\phantom{\ }consecutive\phantom{\ }low-fidelity\phantom{\ }time\phantom{\ }stepping\phantom{\ }blocks}\\\hspace*{0.333em}\hspace*{0.333em}\hspace*{0.333em}\hspace*{0.333em}\hspace*{0.333em}\hspace*{0.333em}$\mathbf{q}^{(p)} \ p = 0,1, \ldots , n-1 \quad$~\texttt{\textbackslash{}\textbackslash{}\phantom{\ }sources\phantom{\ }corresponding\phantom{\ }to\phantom{\ }different\phantom{\ }training\phantom{\ }pairs}\\\hspace*{0.333em}\hspace*{0.333em}\hspace*{0.333em}\hspace*{0.333em}\hspace*{0.333em}\hspace*{0.333em}$\mathbf{u}_{\tau_i}^{(p)}, \ p = 0,1, \ldots , n-1, \ i = 0, 1, \ldots , \lfloor \frac{M-1}{k} \rfloor \quad$~\texttt{\textbackslash{}\textbackslash{}\phantom{\ }high-fidelity\phantom{\ }snapshots}\\\hspace*{0.333em}\hspace*{0.333em}\hspace*{0.333em}\hspace*{0.333em}\hspace*{0.333em}\hspace*{0.333em}${\theta^0_i}, \ i = 0, 1, \ldots , \lfloor \frac{M-1}{k} \rfloor \quad$~\texttt{//\phantom{\ }randomly\phantom{\ }initialized\phantom{\ }parameters}\\\hspace*{0.333em}\hspace*{0.333em}2.~$ItrNum \leftarrow 0$\\\hspace*{0.333em}\hspace*{0.333em}3.~\textbf{FOR}~$i = 0:\lfloor \frac{M-1}{k} \rfloor$~\textbf{DO}\\\hspace*{0.333em}\hspace*{0.333em}4.~~~~~~$\theta_i = \theta^0_i$\\\hspace*{0.333em}\hspace*{0.333em}5.~\textbf{FOR}~$p = 0:n-1$~\textbf{DO}\\\hspace*{0.333em}\hspace*{0.333em}6.~~~~~~$\mathbf{\hat{u}}_0^{(p)} = \bar{F}_k(\mathbf{q}^{(p)})$\\\hspace*{0.333em}\hspace*{0.333em}7.~\textbf{WHILE}~$itrNum < MaxItr$~\textbf{DO}\\\hspace*{0.333em}\hspace*{0.333em}8.~~~~~~\textbf{FOR}~$i = 0:\lfloor \frac{M-1}{k} \rfloor$~\textbf{DO}\\\hspace*{0.333em}\hspace*{0.333em}9.~~~~~~~~~~\textbf{IF}~$i>0$~\textbf{DO}\\\hspace*{0.333em}\hspace*{0.333em}10.~~~~~~~~~~~~~\textbf{FOR}~$p = 0:n-1$~\textbf{DO}\\\hspace*{0.333em}\hspace*{0.333em}11.~~~~~~~~~~~~~~~~~$\mathbf{\hat{u}}_i^{(p)} = \bar{F}_k(\mathcal{G}_{\theta_{i-1}}(\mathbf{\hat{u}}_{i-1}^{(p)}))$\\\hspace*{0.333em}\hspace*{0.333em}12.~~~~~~~~~\textbf{FOR}~$miniItrNum = 1:MaxMiniItr$~\textbf{DO}\\\hspace*{0.333em}\hspace*{0.333em}13.~~~~~~~~~~~~~$p \leftarrow SampleWithoutReplacement(\left \{ 0,1, \ldots , n-1 \right \})$\\\hspace*{0.333em}\hspace*{0.333em}14.~~~~~~~~~~~~~$\theta_i \leftarrow \mathop{\arg \min}_{\theta_i} \left \| \mathcal{G}_{\theta_i}(\mathbf{\hat{u}}_i^{(p)}) - \mathbf{u}_{\tau_i}^{(p)} \right \|_1$\\\hspace*{0.333em}\hspace*{0.333em}15.~~~~~~~~~~~~~$ItrNum \leftarrow ItrNum + 1$\\\hspace*{0.333em}\hspace*{0.333em}16.~\textbf{RETURN}~${\theta_i}, \ i = 0, 1, \ldots , \lfloor \frac{M-1}{k} \rfloor$
\caption{Heuristic for optimizing CNNs
$\mathcal{G}_{\theta_i}, \ i = 0, 1, \ldots , \lfloor \frac{M-1}{k} \rfloor$.}\label{alg:alg1}
\end{scholmdAlgorithm}

\subsection{CNN architecture}\label{cnn-architecture}

Motivated by our previous attempts for numerical dispersion removal from
wavefield snapshots \citep{siahkoohi2018deep, siahkoohi2019transfer}, we
use the exact architecture provided by \citet{johnson2016perceptual},
which includes Residual Blocks, the main building block of ResNets,
introduced by \citet{he2016deep}, for all the CNNs
$\mathcal{G}_{\theta_i}, \ i = 0, 1, \ldots , \lfloor \frac{M-1}{k} \rfloor$.

\subsection{Training details and
implementation}\label{training-details-and-implementation}

While CNNs are known to generalize well---i.e., maintain the quality of
performance when applied to unseen data, they can only be successfully
applied to a data set drawn from the same distribution as the training
data. Because of the Earth's heterogeneity and complex geological
structures present in realistic-looking models, training a neural
network that can generalize well when applied to another velocity model
can become challenging. While we have successfully demonstrated that
transfer learning \citep{yosinski2014transferable} can be used in
situations where the neural network is initially trained on data from a
proximal survey \citep{siahkoohi2019transfer}, we chose in this
contribution, as a proof of concept, to keep the velocity model fixed,
and vary the source locations to generate different training/testing
pairs.

We use the Marmousi velocity model and out of $401$ available shot
locations with $7.5$ $\mathrm{m}$ spacing, we allocate half of the shot
locations to training and use the rest of the shot locations to generate
testing pairs, for evaluation purposes. The maximum simulation time in
our experiments in $1.1$ $\mathrm{s}$.

We designed and implemented our deep architectures in
TensorFlow\footnote{\url{https://www.tensorflow.org/}}. To carry out our
wave-equation simulations with finite differences, we used
Devito\footnote{\url{https://www.devitoproject.org/}}
\citep{devito-api, devito-compiler}. We used the functionality of
Operator Discretization Library\footnote{\url{https://odlgroup.github.io/odl/}}
to wrap Devito operators into a TensorFlow layers. Our implementation
can be found on GitHub\footnote{\url{https://github.com/alisiahkoohi/NN-augmented-wave-sim}}.

We ran our algorithm on Amazon Web Services' g3.4xlarge instance, where
we optimize the CNN parameters on a NVIDIA Tesla M60 GPU and Devito
utilizes $16$ CPU cores to perform finite-difference wave-equation
simulations. Initially, we simulate the high-fidelity training wavefield
snapshots,
$\mathbf{u}_{\tau_i}^{(p)}, \ p = 0,1, \ldots , n-1, \ i = 0, 1, \ldots , \lfloor \frac{M-1}{k} \rfloor$,
only once, in the beginning, and store them. In order to limit CPU-GPU
communication, before utilizing the GPU to to update
${\theta_i}, \ i = 0, 1, \ldots , \lfloor \frac{M-1}{k} \rfloor$, we
generate the input to $i^{\text{th}}$ CNN,
$\mathbf{\hat{u}}_i^{(p)}, \ p = 0,1, \ldots , n-1$ all at once, and
store them. Afterwards, $i^{\text{th}}$ CNN can be (re)trained using the
stored input/output wavefield snapshot pairs for several
mini-iterations.

\section{Numerical experiments}\label{numerical-experiments}

We want to indicate that neural networks, when augmented with inaccurate
physics, e.g., a poor discretization of Laplacian, are able to
approximate the wavefields obtained by an accurate approximation to wave
equation. To demonstrate this, we conduct three numerical experiments in
which we keep the velocity model fixed, and vary the source locations to
generate different training/testing pairs. The experiments differ in the
number of CCNs used throughout learned wave propagation. We use three,
five, and ten CNNs while keeping the total number of iterations fixed.
This implies that an experiment with more CNNs, optimizes each CNN with
a smaller number of iterations per CNN, because,
$\text{iterations per CNN} \times \text{number of CNNs} = \text{total number of iterations}$.

A neural network augmented wave simulator with $n_1$ CNNs needs more
training iterations and possibly more training data to perform equally
as well as a neural network augmented wave simulator with $n_2$ CNNs,
when $n_1 > n_2$. For a fixed number of total iterations, iterations per
CNN is inversely proportional to number of CNNs utilized. Therefore, the
first $n_2$ CNNs in the neural network augmented wave simulator with
$n_1$ CNNs will perform worse than the CNNs in the wave propagator with
$n_2$ CNNs. Consequently, the error accumulated by the poor performance
of first $n_2$ CNNs, combined with artifacts introduced by low-fidelity
timestepping makes the matters worse for the later CNNs in the more
complex learned wave propagator. Therefore, in our experiments, since
the total number of iterations is fixed, we expect to see the quality of
dispersion removal degrade as the number of CNNs increase in a learned
wave propagator. Table~\ref{training-details} summarizes the total
number of iterations, iterations per CNN, training pairs, training time,
and number of tunable parameters for the three different experiments.

\begin{table}
\centering
\begin{tabular}{cccccc}
\toprule\addlinespace
CNNs & Iterations & Iterations per CNN & Pairs per CNN & Time & Param.
count\tabularnewline
\midrule
$3$ & $100500$ & $33500$ & $201$ & $17.99$ hours &
$34150272$\tabularnewline
$5$ & $100500$ & $20100$ & $201$ & $19.79$ hours &
$56917120$\tabularnewline
$10$ & $100500$ & $10050$ & $201$ & $49.24$ hours &
$113834240$\tabularnewline
\bottomrule
\end{tabular}
\caption{Summary of details in the three neural network augmented
wave-equation simulation experiments.}\label{training-details}
\end{table}

\begin{figure}
\centering
\subfloat[\label{corr-3-snr-0}]{\includegraphics[width=0.500\hsize]{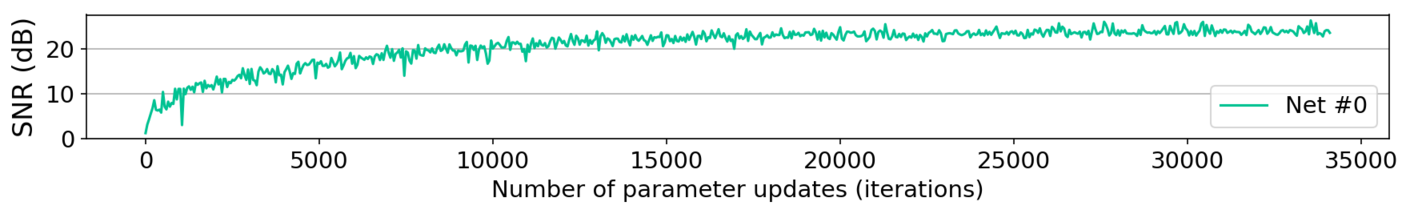}}
\subfloat[\label{corr-3-loss-0}]{\includegraphics[width=0.500\hsize]{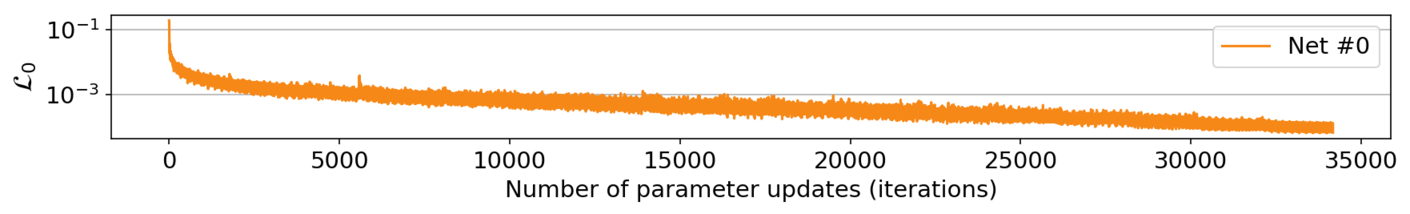}}
\\
\subfloat[\label{corr-3-snr-1}]{\includegraphics[width=0.500\hsize]{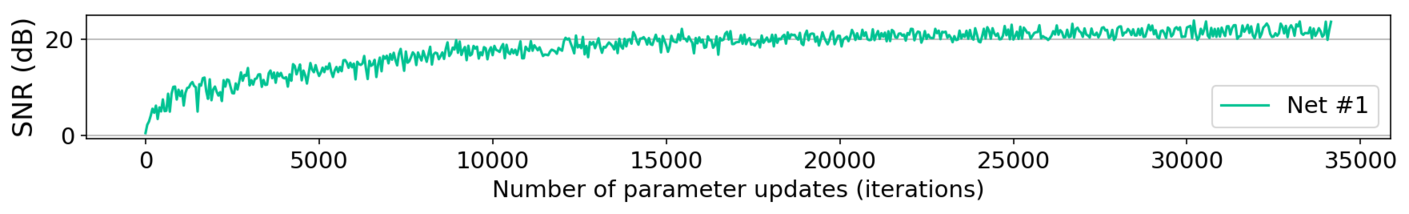}}
\subfloat[\label{corr-3-loss-1}]{\includegraphics[width=0.500\hsize]{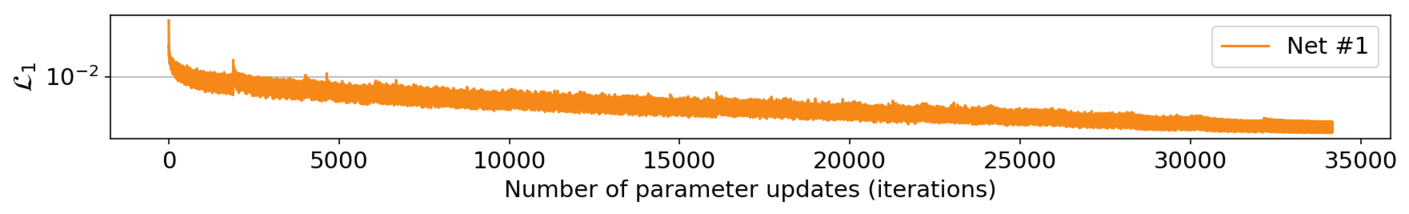}}
\\
\subfloat[\label{corr-3-snr-2}]{\includegraphics[width=0.500\hsize]{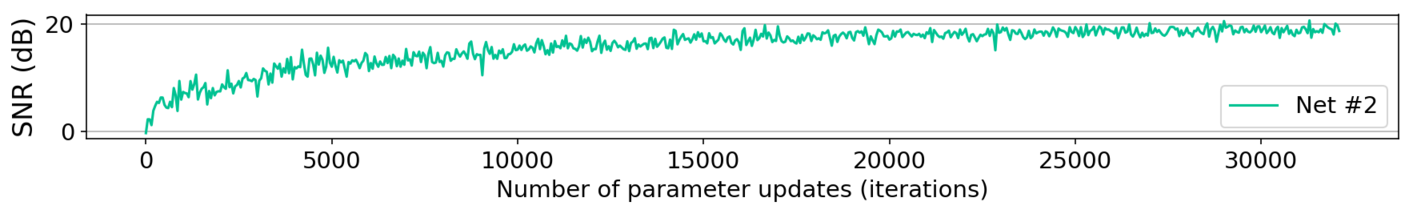}}
\subfloat[\label{corr-3-loss-2}]{\includegraphics[width=0.500\hsize]{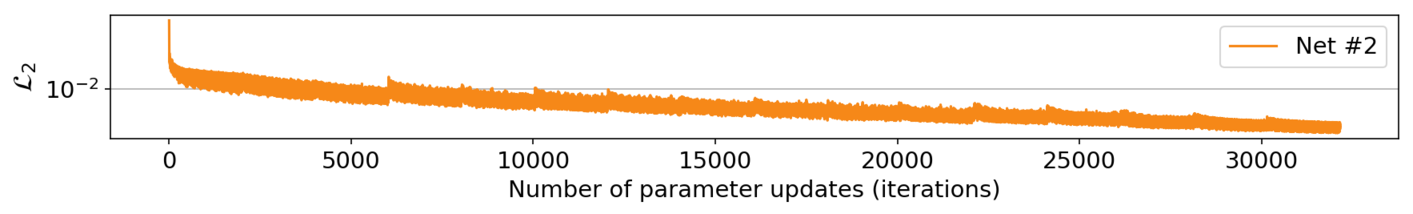}}
\caption{Neural network augmented wave simulation with three CNNs. First
column from top to bottom: SNR curves, evaluated on testing pairs during
training, for a) the first to e) the last CNN, in order. Second column
from top to bottom: training objective function value curves, evaluated
on training pairs, for b) the first to the f) last CNN, in
order.}\label{corr-3-log}
\end{figure}

\begin{figure}
\centering
\subfloat[\label{corr-5-snr-0}]{\includegraphics[width=0.500\hsize]{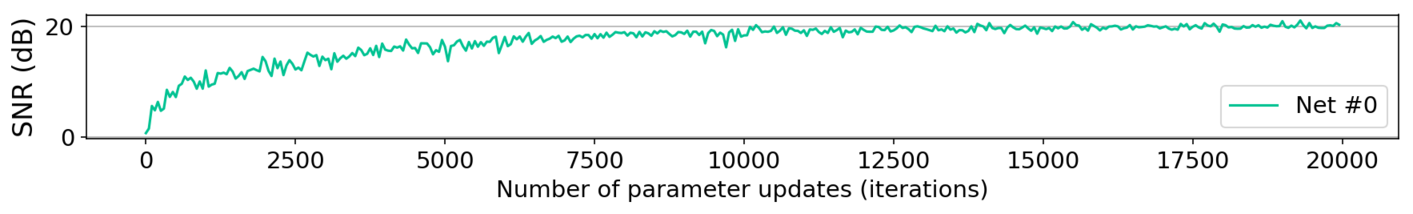}}
\subfloat[\label{corr-5-loss-0}]{\includegraphics[width=0.500\hsize]{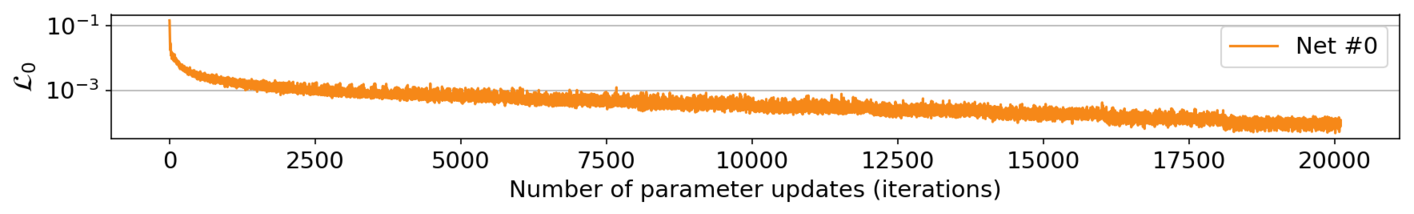}}
\\
\subfloat[\label{corr-5-snr-1}]{\includegraphics[width=0.500\hsize]{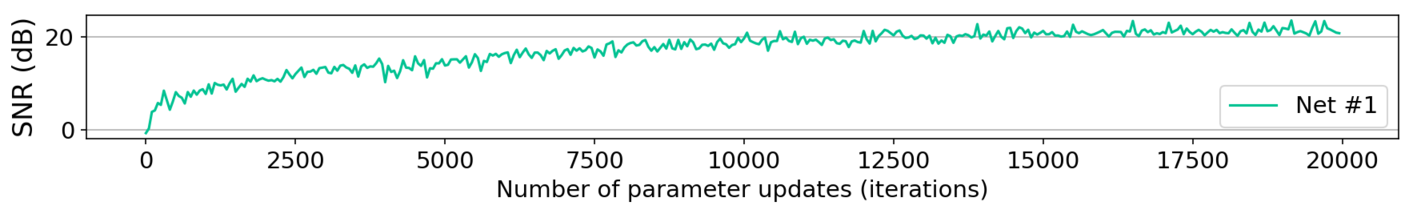}}
\subfloat[\label{corr-5-loss-1}]{\includegraphics[width=0.500\hsize]{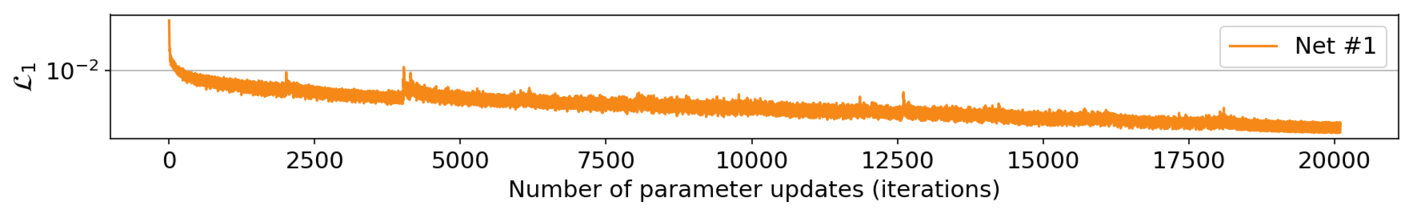}}
\\
\subfloat[\label{corr-5-snr-2}]{\includegraphics[width=0.500\hsize]{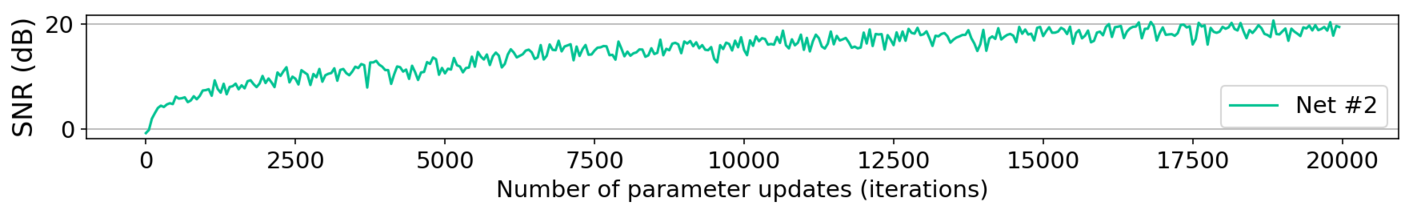}}
\subfloat[\label{corr-5-loss-2}]{\includegraphics[width=0.500\hsize]{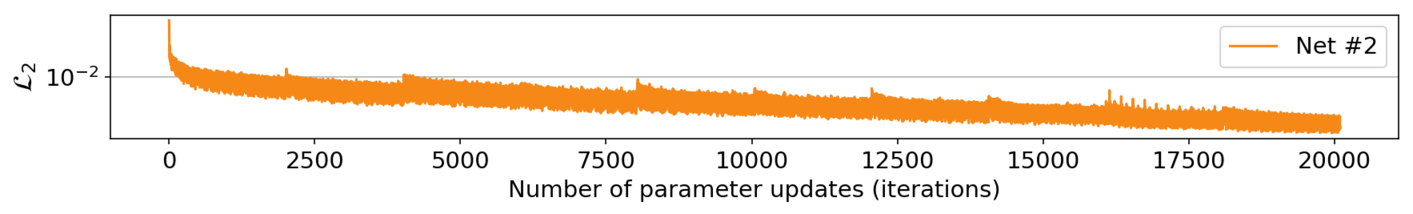}}
\\
\subfloat[\label{corr-5-snr-3}]{\includegraphics[width=0.500\hsize]{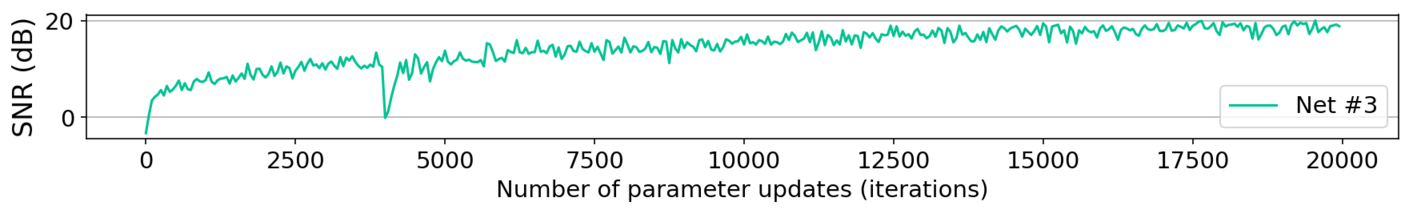}}
\subfloat[\label{corr-5-loss-3}]{\includegraphics[width=0.500\hsize]{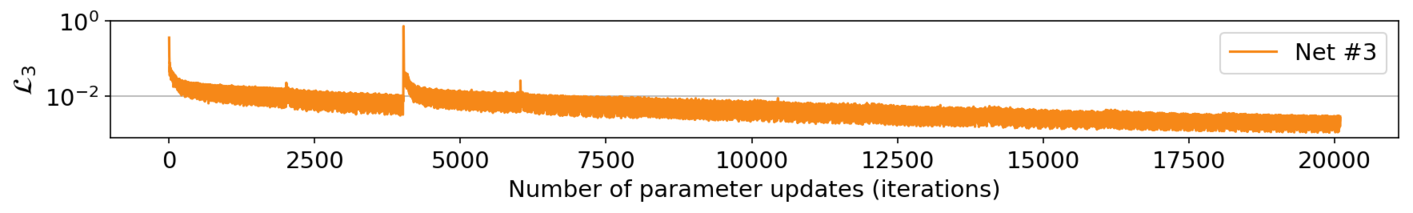}}
\\
\subfloat[\label{corr-5-snr-4}]{\includegraphics[width=0.500\hsize]{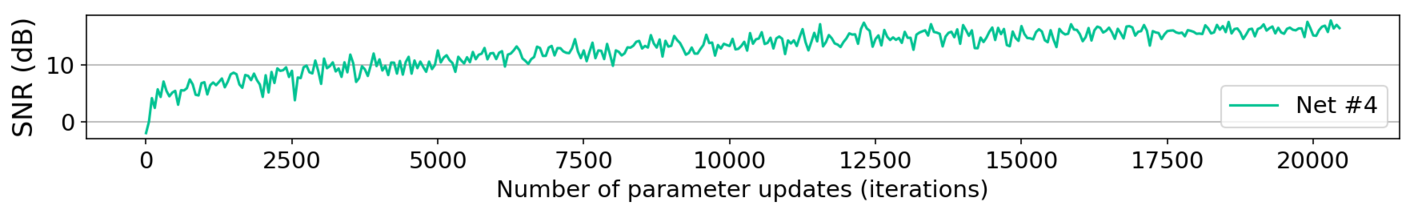}}
\subfloat[\label{corr-5-loss-4}]{\includegraphics[width=0.500\hsize]{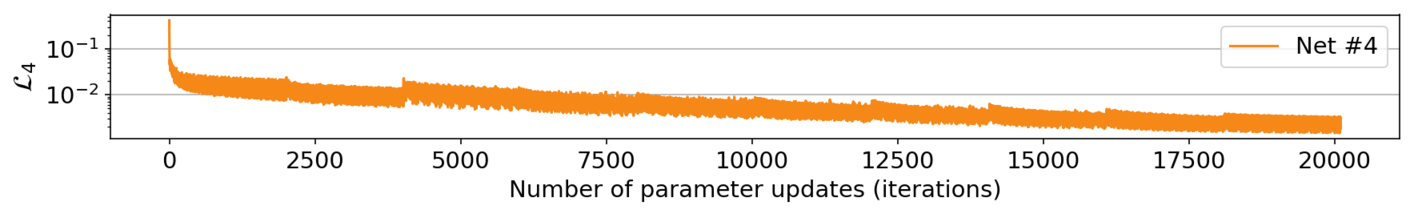}}
\caption{Neural network augmented wave simulation with five CNNs. First
column from top to bottom: SNR curves, evaluated on testing pairs during
training, for a) the first to i) the last CNN, in order. Second column
from top to bottom: training objective function value curves, evaluated
on training pairs, for b) the first to the j) last CNN, in
order.}\label{corr-5-log}
\end{figure}

\begin{figure}
\centering
\subfloat[\label{corr-10-snr-0}]{\includegraphics[width=0.500\hsize]{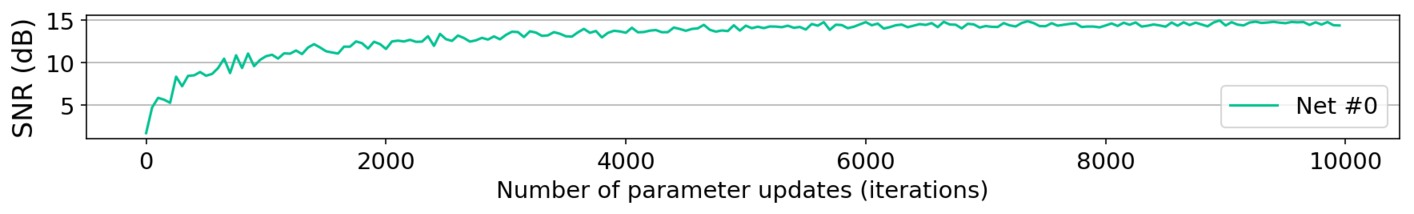}}
\subfloat[\label{corr-10-loss-0}]{\includegraphics[width=0.500\hsize]{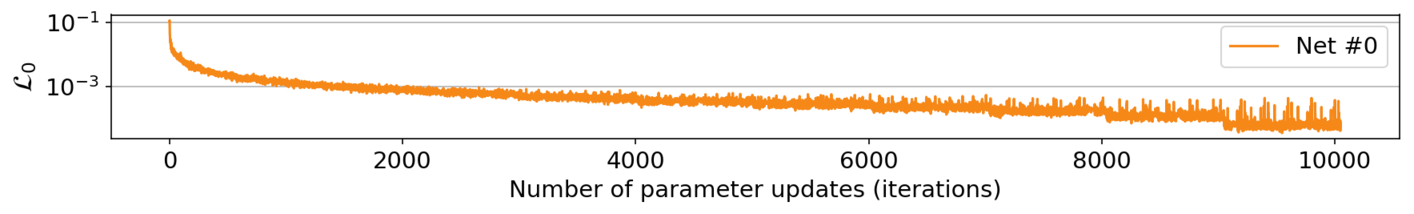}}
\\
\subfloat[\label{corr-10-snr-1}]{\includegraphics[width=0.500\hsize]{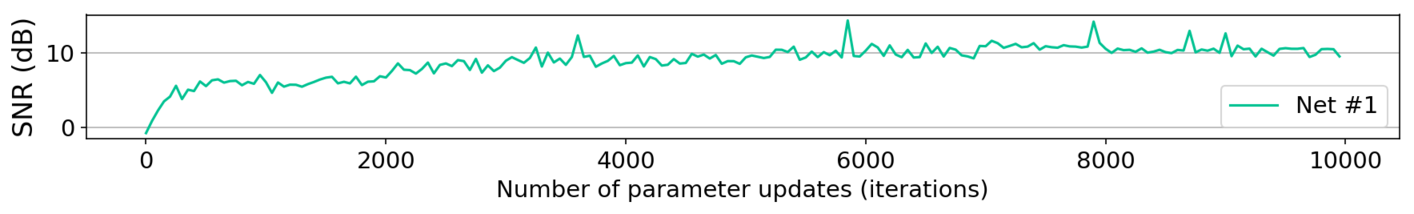}}
\subfloat[\label{corr-10-loss-1}]{\includegraphics[width=0.500\hsize]{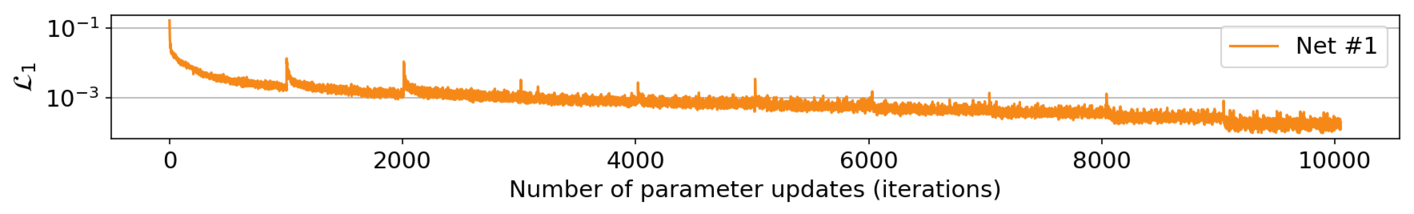}}
\\
\subfloat[\label{corr-10-snr-2}]{\includegraphics[width=0.500\hsize]{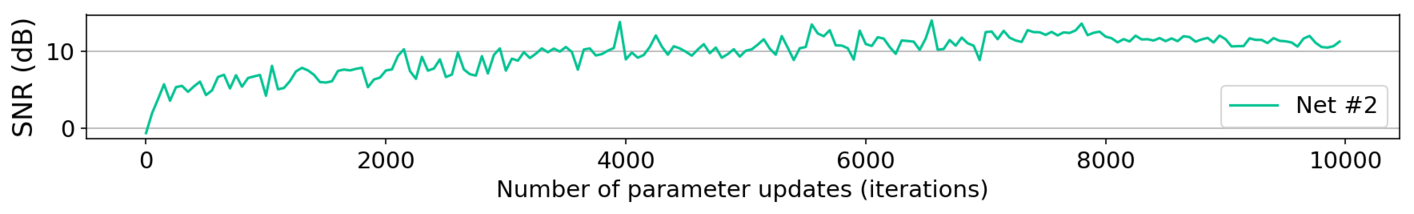}}
\subfloat[\label{corr-10-loss-2}]{\includegraphics[width=0.500\hsize]{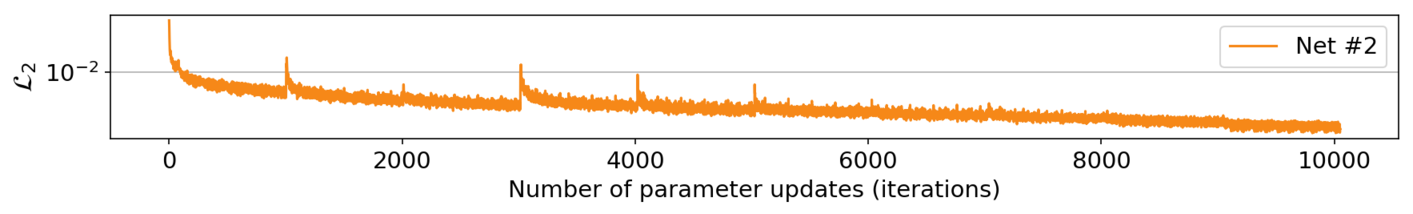}}
\\
\subfloat[\label{corr-10-snr-3}]{\includegraphics[width=0.500\hsize]{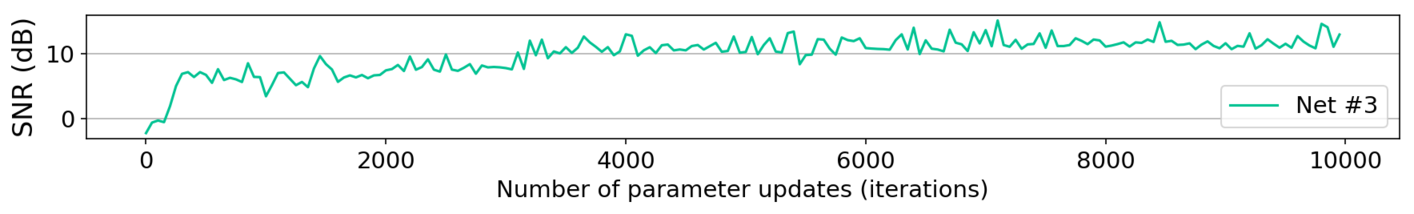}}
\subfloat[\label{corr-10-loss-3}]{\includegraphics[width=0.500\hsize]{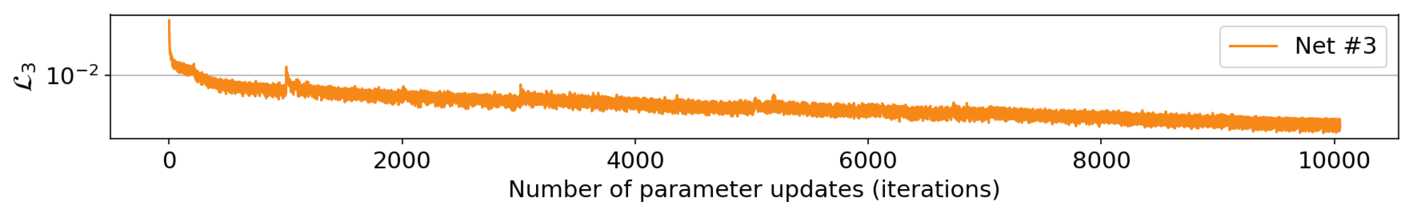}}
\\
\subfloat[\label{corr-10-snr-4}]{\includegraphics[width=0.500\hsize]{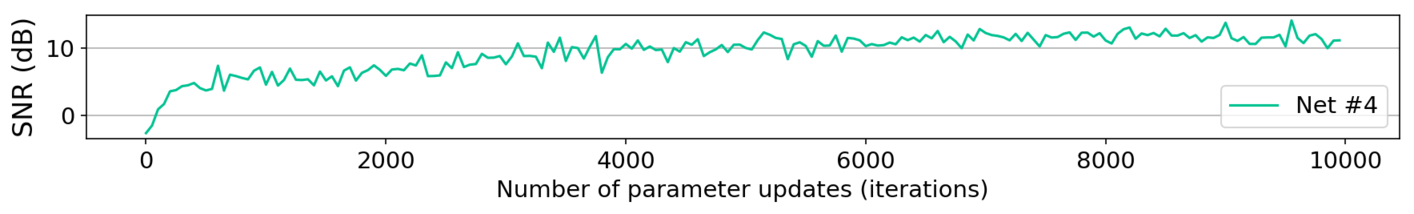}}
\subfloat[\label{corr-10-loss-4}]{\includegraphics[width=0.500\hsize]{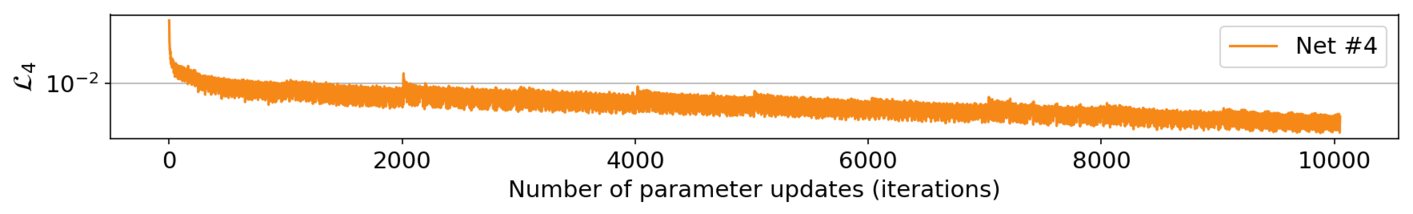}}
\\
\subfloat[\label{corr-10-snr-5}]{\includegraphics[width=0.500\hsize]{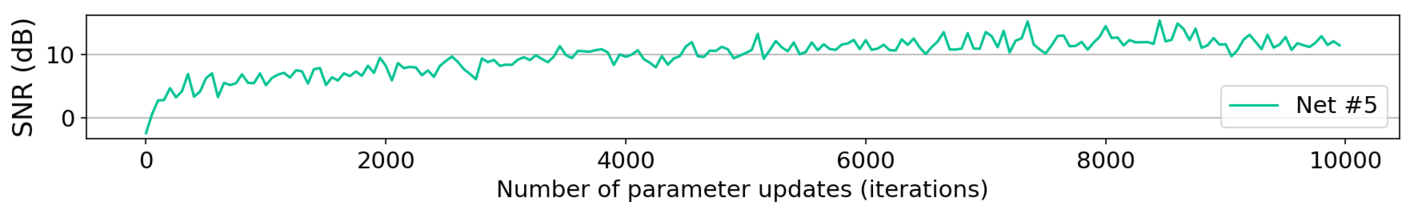}}
\subfloat[\label{corr-10-loss-5}]{\includegraphics[width=0.500\hsize]{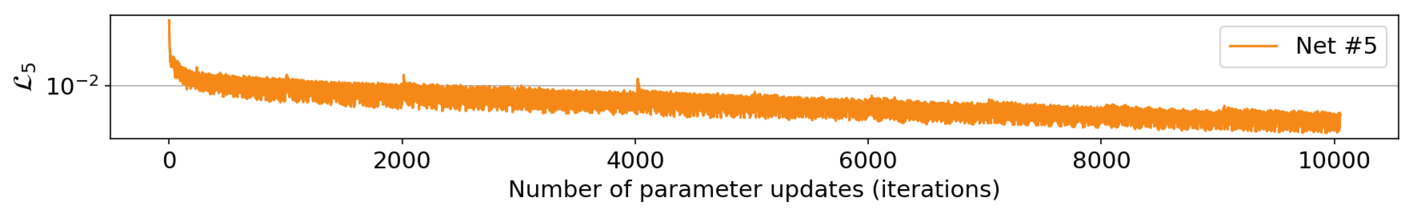}}
\\
\subfloat[\label{corr-10-snr-6}]{\includegraphics[width=0.500\hsize]{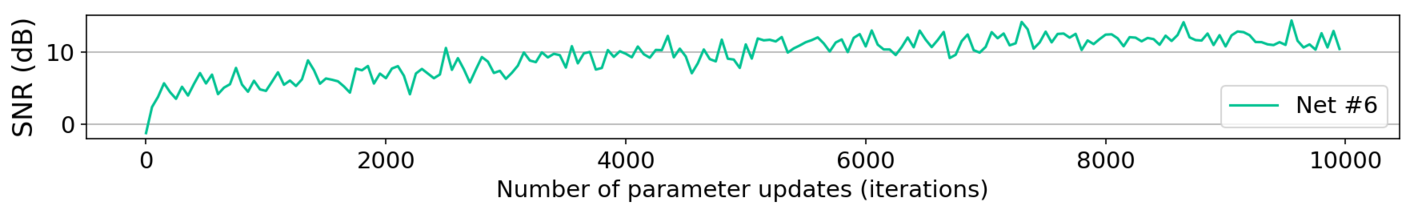}}
\subfloat[\label{corr-10-loss-6}]{\includegraphics[width=0.500\hsize]{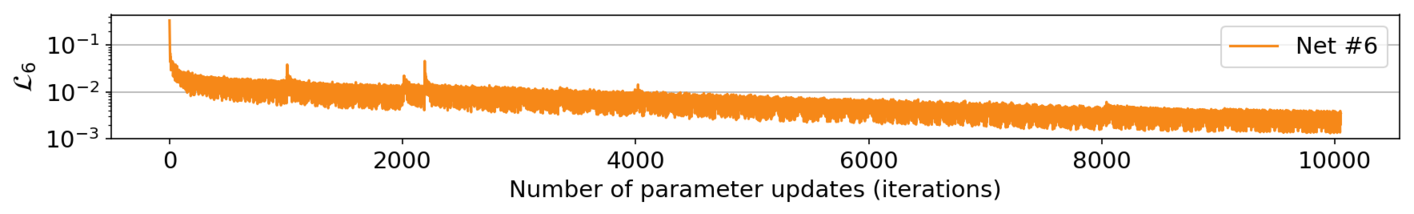}}
\\
\subfloat[\label{corr-10-snr-7}]{\includegraphics[width=0.500\hsize]{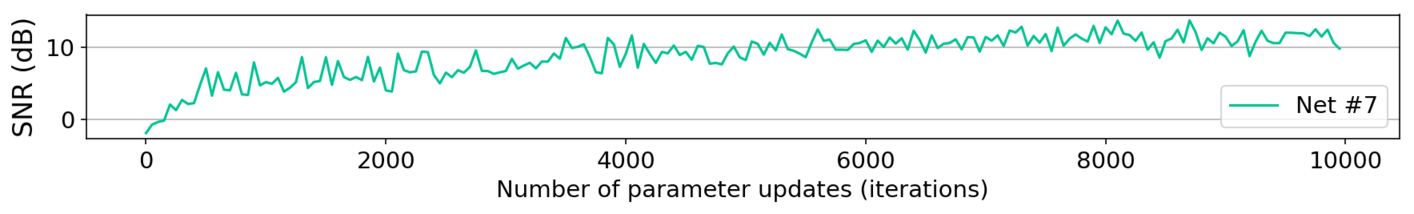}}
\subfloat[\label{corr-10-loss-7}]{\includegraphics[width=0.500\hsize]{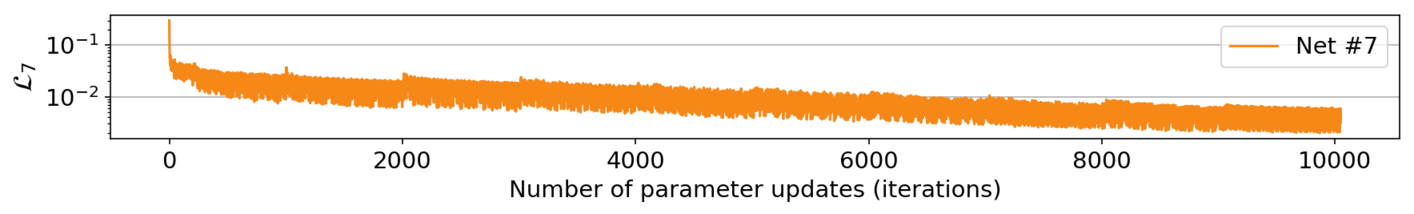}}
\\
\subfloat[\label{corr-10-snr-8}]{\includegraphics[width=0.500\hsize]{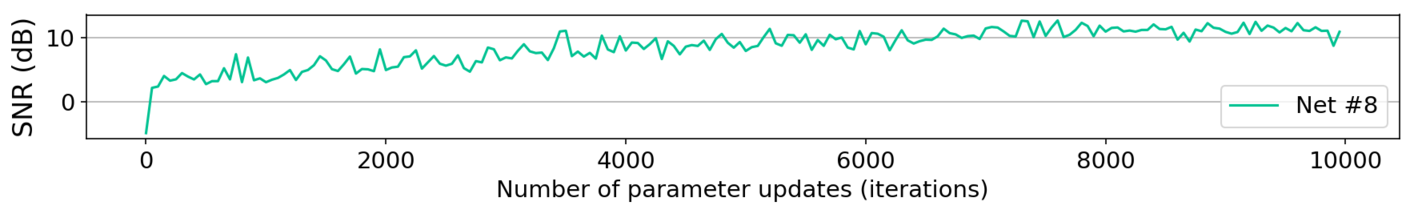}}
\subfloat[\label{corr-10-loss-8}]{\includegraphics[width=0.500\hsize]{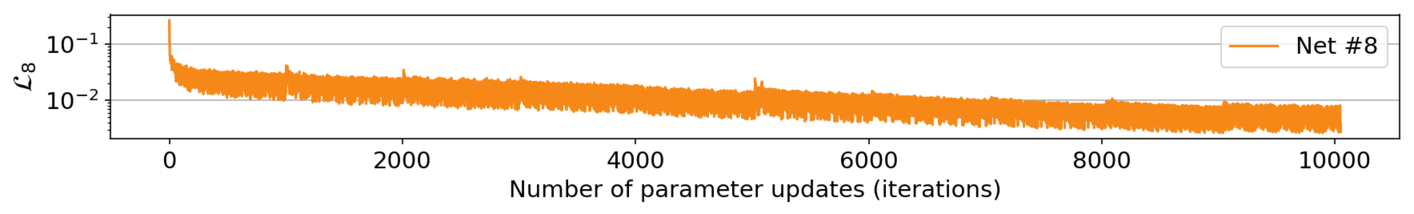}}
\\
\subfloat[\label{corr-10-snr-9}]{\includegraphics[width=0.500\hsize]{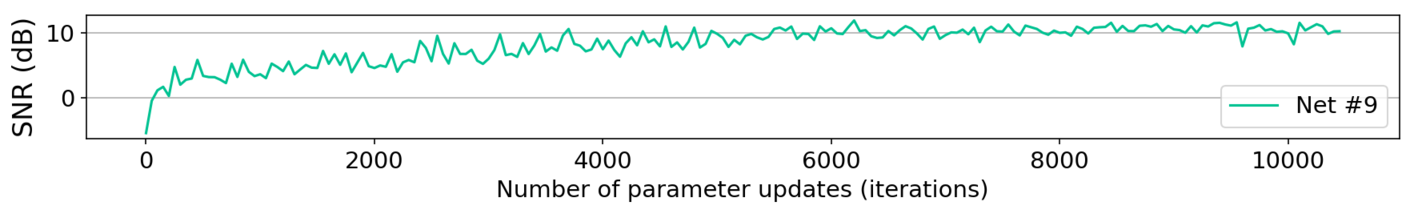}}
\subfloat[\label{corr-10-loss-9}]{\includegraphics[width=0.500\hsize]{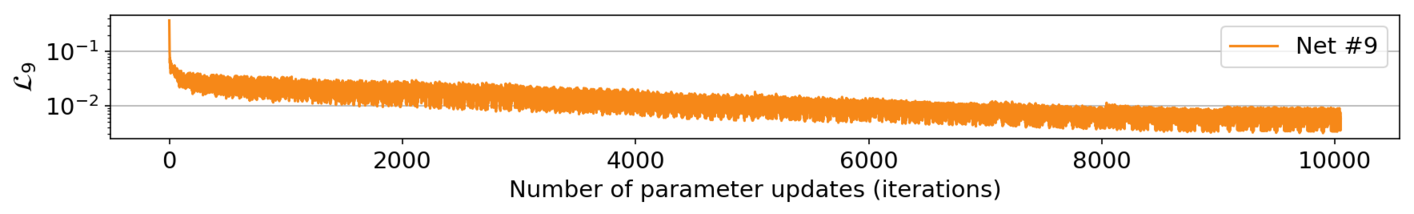}}
\caption{Neural network augmented wave simulation with ten CNNs. First
column from top to bottom: SNR curves, evaluated on testing pairs during
training, for a) the first to s) the last CNN, in order. Second column
from top to bottom: training objective function value curves, evaluated
on training pairs, for b) the first to the t) last CNN, in
order.}\label{corr-10-log}
\end{figure}

As described earlier and presented in Table~\ref{training-details}, the
\texttt{MaxItr} variable used in the \texttt{While} condition in line
$8$ of Algorithm~\ref{alg:alg1} is set to $500$ for all our experiments.
Figures~\ref{corr-3-log} $-$ \ref{corr-10-log} demonstrate the values of
objective function presented in Equation~\ref{trainCNN-i} in orange, and
the wavefield correction signal-to-noise ratio (SNR) in blue, evaluated
on testing data pairs during training, for experiments with three, five,
and ten CNNs, respectively. Note that the SNR curves have not been used
to determine when to stop training and they are only depicted for
demonstration purposes.

Figures~\ref{corr-3-snr-0}, \ref{corr-3-snr-1}, and~\ref{corr-3-snr-2}
show the wavefield correction SNR for first, second, and third CNN,
respectively, in the neural network augmented wave simulator that
includes three CNNs. Similarly, Figures~\ref{corr-3-loss-0},
\ref{corr-3-loss-1}, and~\ref{corr-3-loss-2} depict the training
objective values throughout training for first, second, and third CNNs.
As it can be seen from objective function curves, the raining heuristic
has been effective and the objective function values have a decreasing
trend. Note that CNNs has been trained for $33500$ iterations, on
average, with a total of $100500$ iterations. Several equispaced spikes
can be noticed on the objective function value curves. For instance, see
the objective value function curve of the third CNN, in
Figure~\ref{corr-3-loss-2}, at $6030, 8040, 10050, \text{and } 12060$
iterations. The mini-batch we use in this experiment is $10$. Those
spikes occur in moments in training when we have started retraining the
third CNN, after updating the first and second CNNs. As discussed
before, a change in the parameters of the CNNs preceding a CNN causes
changes in the input of the later CNN, and consequently the objective
function becomes large when starting to retrain the CNNs in later stages
again.

Similar objective function value and SNR curves for other two neural
network augmented wave propagators, utilizing five and ten CNNs, can be
found in Figures~\ref{corr-5-log} and~\ref{corr-10-log}, respectively.
First column in Figures~\ref{corr-5-log} and~\ref{corr-10-log}, from top
to bottom, indicate SNR of wavefield correction obtained by the first to
the last CNN, evaluated on testing pairs while training, respectively.
The second column of Figures~\ref{corr-5-log} and~\ref{corr-10-log}
indicate the objective function value curves throughout optimization of
Equation~\ref{trainCNN-i} for training neural network augmented wave
propagators, utilizing five and ten CNNs, respectively. In both columns,
from top to bottom, the objective function value curves correspond to
CNNs from beginning to the end of the learned wave propagators, in
order.

We make two main observations from Figures~\ref{corr-5-log}
and~\ref{corr-10-log}. First, the objective function values indicate
overall decreases, validating effectiveness of the introduced heuristic.
Also, the spikes on the objective function value curve can be seen,
which are correlated with the stages in training when
Algorithm~\ref{alg:alg1} revisits a CNN after updating the rest of the
CNN parameters. As explained before, spikes are caused by change in
parameters of preceding CNNs to a CNN, which in turn alters the input
training wavefields of the CNN. Second, due to decrease in iterations
per CNN as the number of CNNs increases, the SNR curves converge to a
lower value when the number of CNNs increase.

\begin{figure}
\centering
\subfloat[\label{corr-3-true-0}]{\includegraphics[width=0.250\hsize]{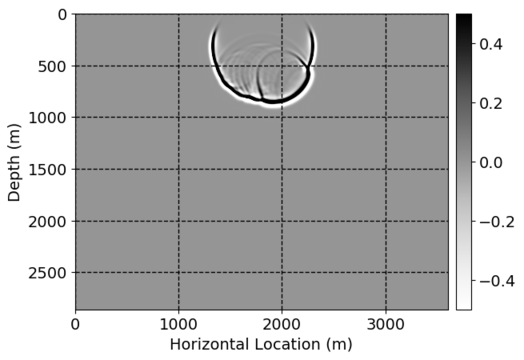}}
\subfloat[\label{corr-3-dsp-0}]{\includegraphics[width=0.250\hsize]{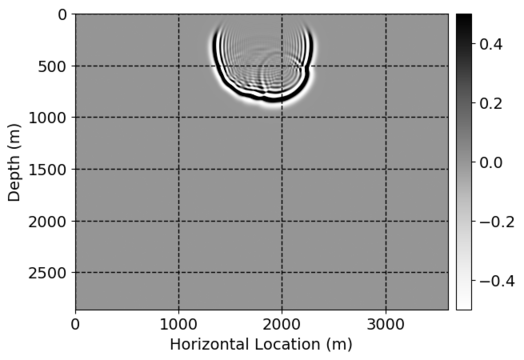}}
\subfloat[\label{corr-3-res-0}]{\includegraphics[width=0.250\hsize]{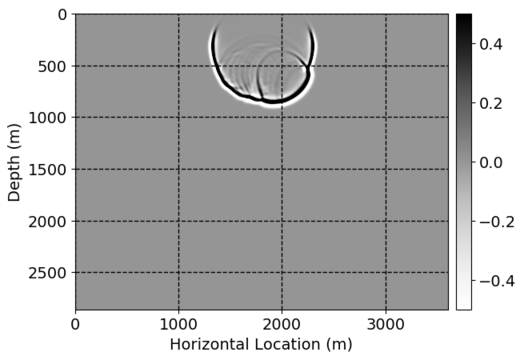}}
\subfloat[\label{corr-3-err-0}]{\includegraphics[width=0.250\hsize]{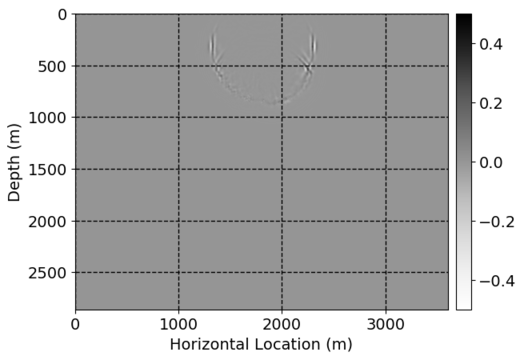}}
\\
\subfloat[\label{corr-3-true-1}]{\includegraphics[width=0.250\hsize]{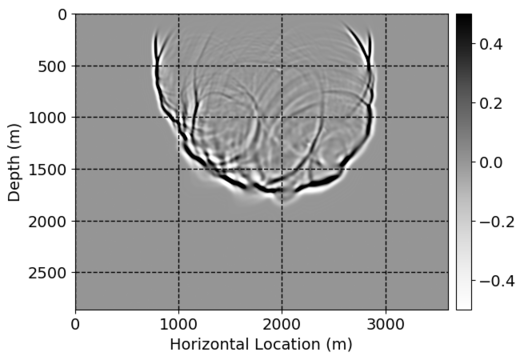}}
\subfloat[\label{corr-3-dsp-1}]{\includegraphics[width=0.250\hsize]{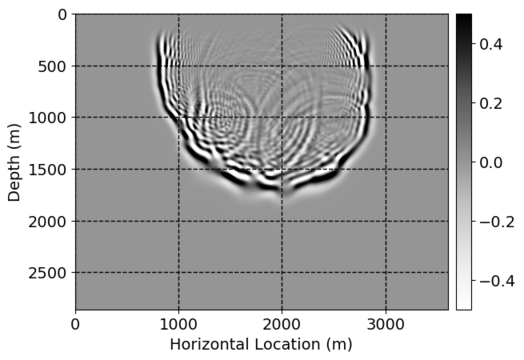}}
\subfloat[\label{corr-3-res-1}]{\includegraphics[width=0.250\hsize]{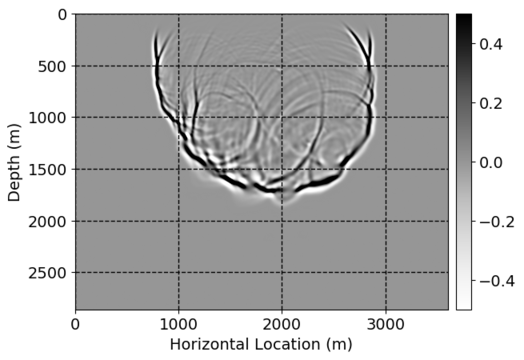}}
\subfloat[\label{corr-3-err-1}]{\includegraphics[width=0.250\hsize]{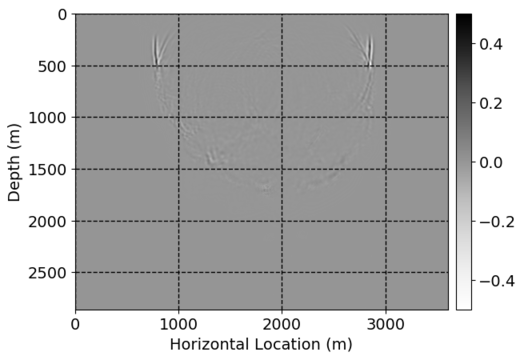}}
\\
\subfloat[\label{corr-3-true-2}]{\includegraphics[width=0.250\hsize]{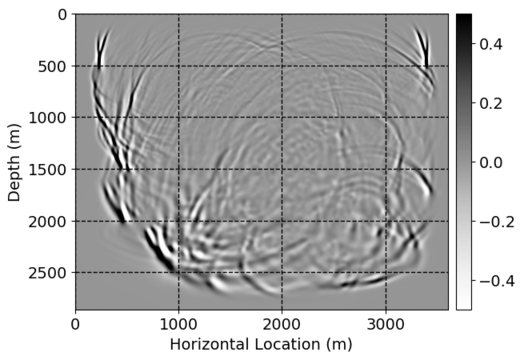}}
\subfloat[\label{corr-3-dsp-2}]{\includegraphics[width=0.250\hsize]{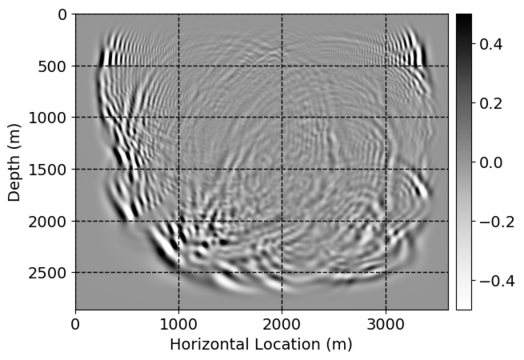}}
\subfloat[\label{corr-3-res-2}]{\includegraphics[width=0.250\hsize]{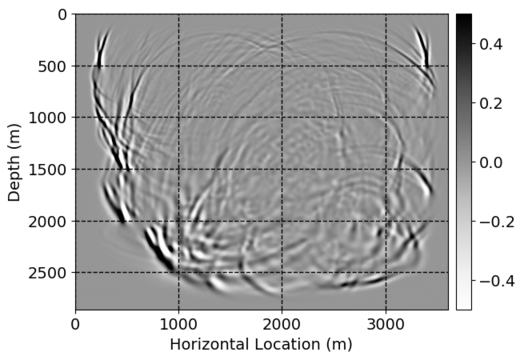}}
\subfloat[\label{corr-3-err-2}]{\includegraphics[width=0.250\hsize]{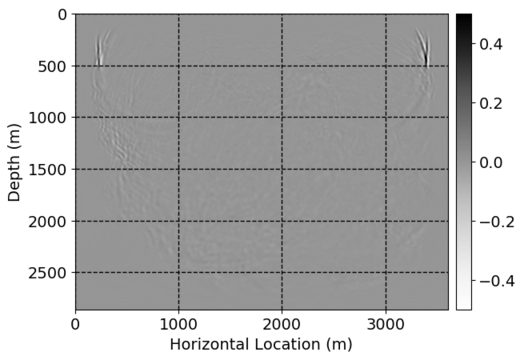}}
\caption{Neural network augmented wave simulation with three CNNs. First
column from top to bottom: a, e, i) high-fidelity wavefield snapshots,
in order. Second column from top to bottom: b, f, j) low-fidelity
wavefield snapshots simulated by solving Equation~\ref{low-fidelity}
with the same simulation time as high-fidelity wavefields, in order.
Third column from top to bottom: c, g, k) result of neural network
augmented wave-equation simulation. Output of the first, second, and the
last CNN, in order. Fourth column from top to bottom: d, h, l)
difference between first and third column, in order.}\label{corr-3}
\end{figure}

\begin{figure}
\centering
\subfloat[\label{corr-5-true-0}]{\includegraphics[width=0.250\hsize]{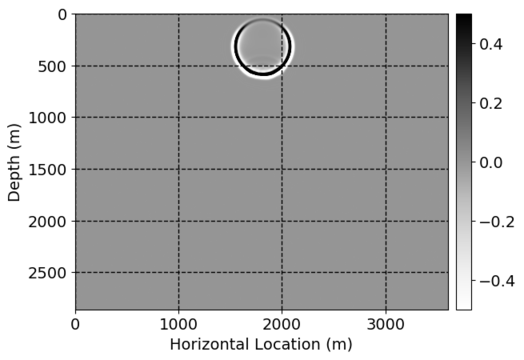}}
\subfloat[\label{corr-5-dsp-0}]{\includegraphics[width=0.250\hsize]{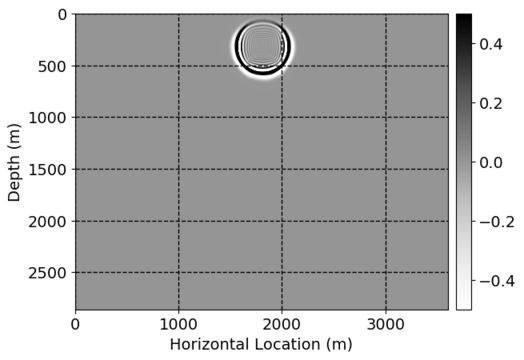}}
\subfloat[\label{corr-5-res-0}]{\includegraphics[width=0.250\hsize]{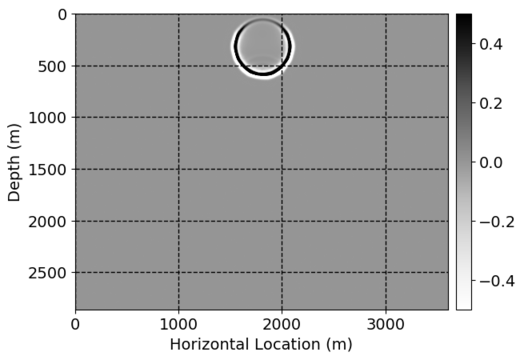}}
\subfloat[\label{corr-5-err-0}]{\includegraphics[width=0.250\hsize]{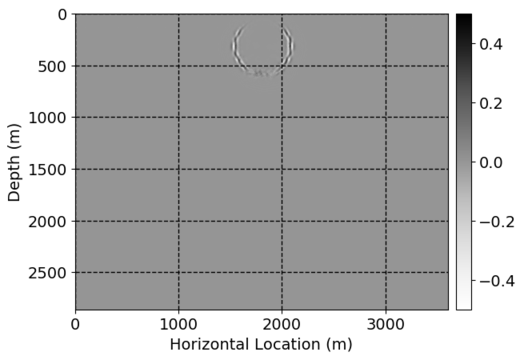}}
\\
\subfloat[\label{corr-5-true-1}]{\includegraphics[width=0.250\hsize]{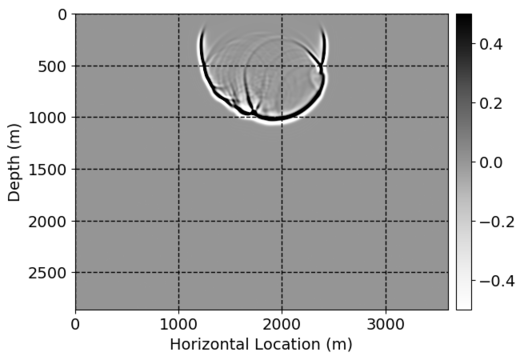}}
\subfloat[\label{corr-5-dsp-1}]{\includegraphics[width=0.250\hsize]{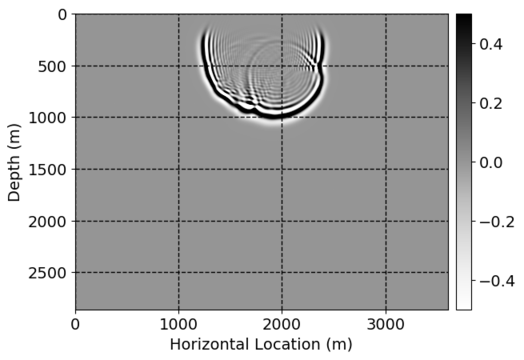}}
\subfloat[\label{corr-5-res-1}]{\includegraphics[width=0.250\hsize]{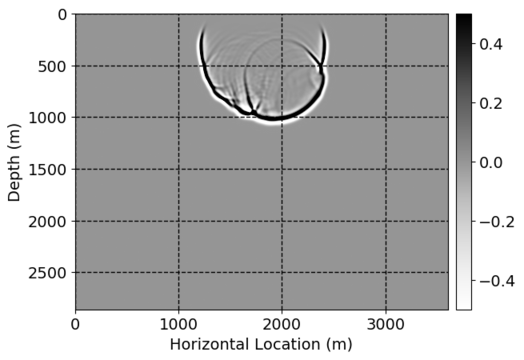}}
\subfloat[\label{corr-5-err-1}]{\includegraphics[width=0.250\hsize]{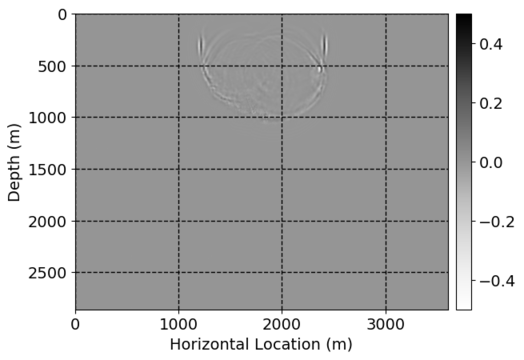}}
\\
\subfloat[\label{corr-5-true-2}]{\includegraphics[width=0.250\hsize]{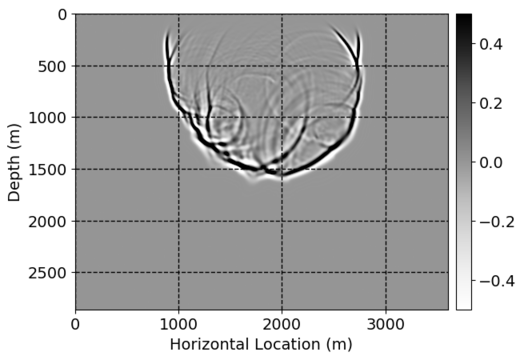}}
\subfloat[\label{corr-5-dsp-2}]{\includegraphics[width=0.250\hsize]{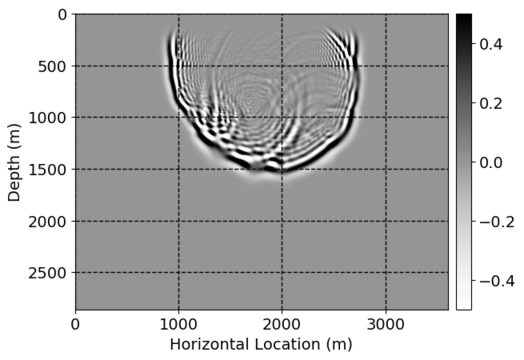}}
\subfloat[\label{corr-5-res-2}]{\includegraphics[width=0.250\hsize]{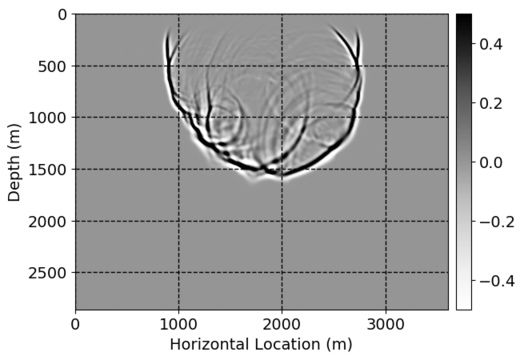}}
\subfloat[\label{corr-5-err-2}]{\includegraphics[width=0.250\hsize]{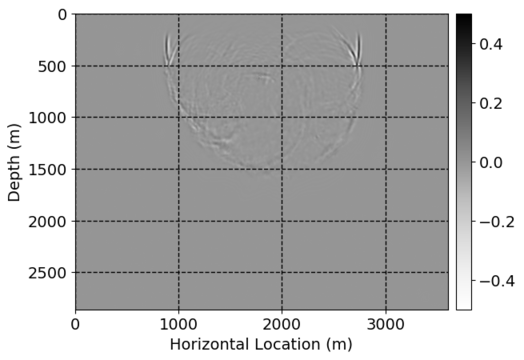}}
\\
\subfloat[\label{corr-5-true-3}]{\includegraphics[width=0.250\hsize]{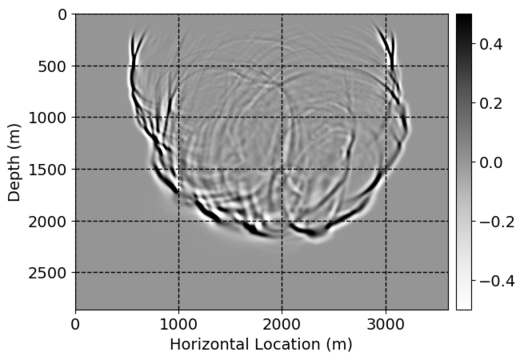}}
\subfloat[\label{corr-5-dsp-3}]{\includegraphics[width=0.250\hsize]{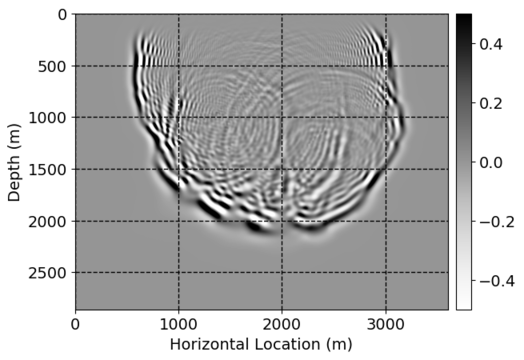}}
\subfloat[\label{corr-5-res-3}]{\includegraphics[width=0.250\hsize]{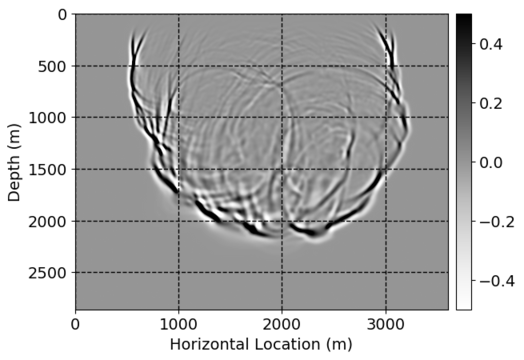}}
\subfloat[\label{corr-5-err-3}]{\includegraphics[width=0.250\hsize]{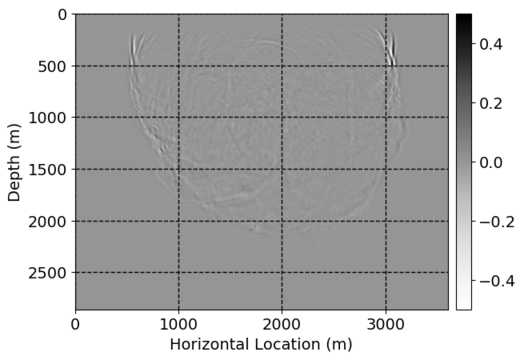}}
\\
\subfloat[\label{corr-5-true-4}]{\includegraphics[width=0.250\hsize]{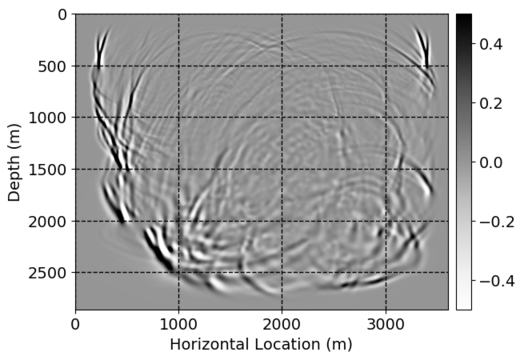}}
\subfloat[\label{corr-5-dsp-4}]{\includegraphics[width=0.250\hsize]{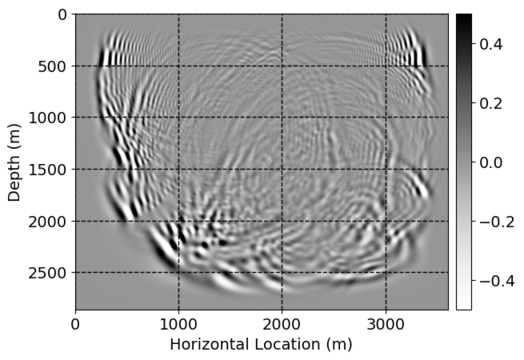}}
\subfloat[\label{corr-5-res-4}]{\includegraphics[width=0.250\hsize]{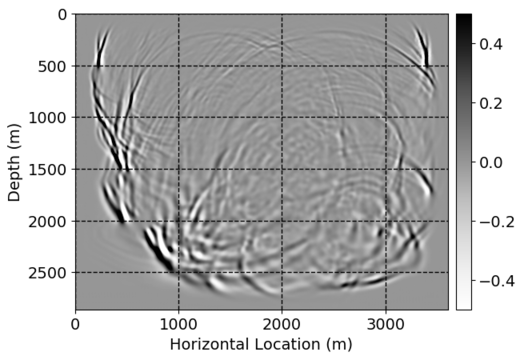}}
\subfloat[\label{corr-5-err-4}]{\includegraphics[width=0.250\hsize]{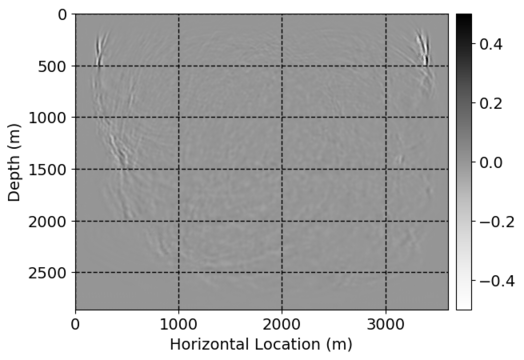}}
\caption{Neural network augmented wave simulation with five CNNs. First
column from top to bottom: a) to q) high-fidelity wavefield snapshots,
in order. Second column from top to bottom: b) to r) low-fidelity
wavefield snapshots simulated by solving Equation~\ref{low-fidelity}
with the same simulation time as high-fidelity wavefields, in order.
Third column from top to bottom: c) to s) result of neural network
augmented wave-equation simulation. Output of the first to the last CNN,
in order. Fourth column from top to bottom: d) to t) difference between
first and third column, in order.}\label{corr-5}
\end{figure}

\begin{figure}
\centering
\subfloat[\label{corr-10-true-0}]{\includegraphics[width=0.250\hsize]{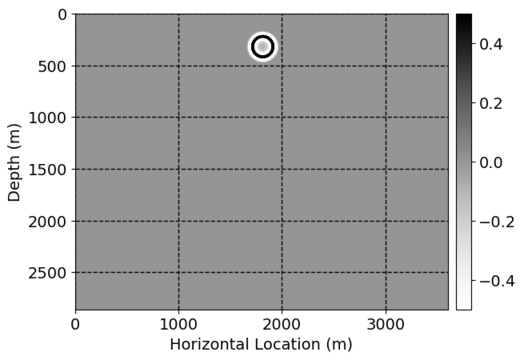}}
\subfloat[\label{corr-10-dsp-0}]{\includegraphics[width=0.250\hsize]{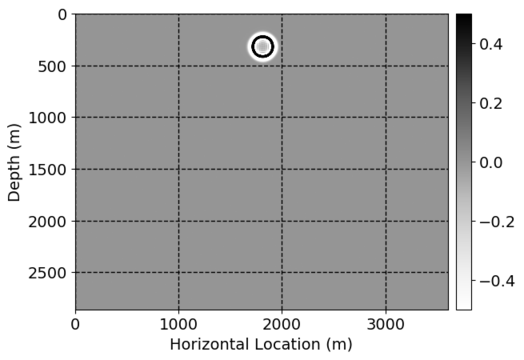}}
\subfloat[\label{corr-10-res-0}]{\includegraphics[width=0.250\hsize]{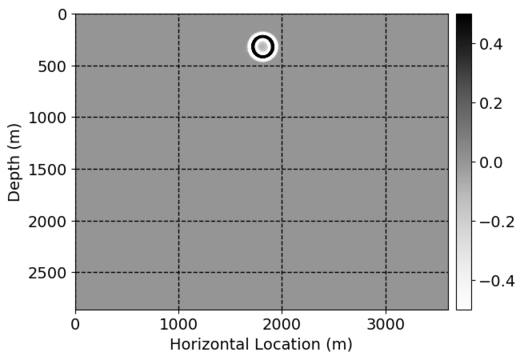}}
\subfloat[\label{corr-10-err-0}]{\includegraphics[width=0.250\hsize]{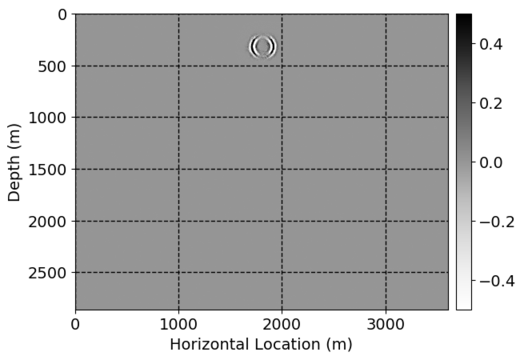}}
\\
\subfloat[\label{corr-10-true-1}]{\includegraphics[width=0.250\hsize]{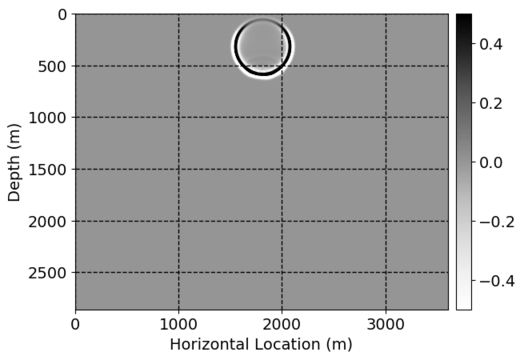}}
\subfloat[\label{corr-10-dsp-1}]{\includegraphics[width=0.250\hsize]{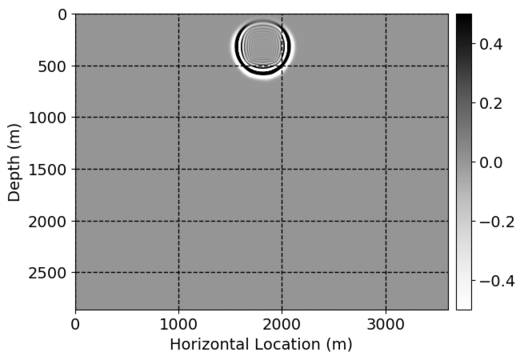}}
\subfloat[\label{corr-10-res-1}]{\includegraphics[width=0.250\hsize]{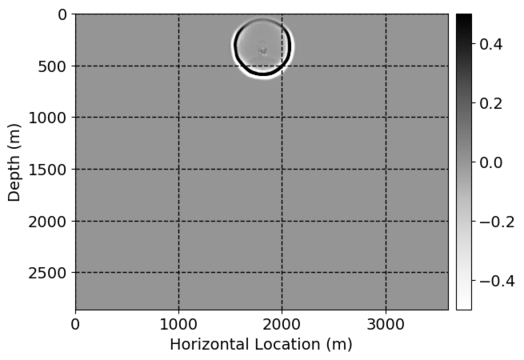}}
\subfloat[\label{corr-10-err-1}]{\includegraphics[width=0.250\hsize]{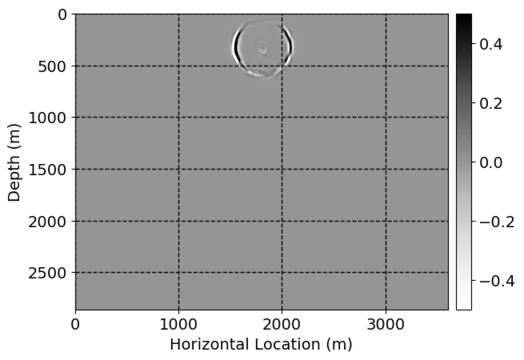}}
\\
\subfloat[\label{corr-10-true-2}]{\includegraphics[width=0.250\hsize]{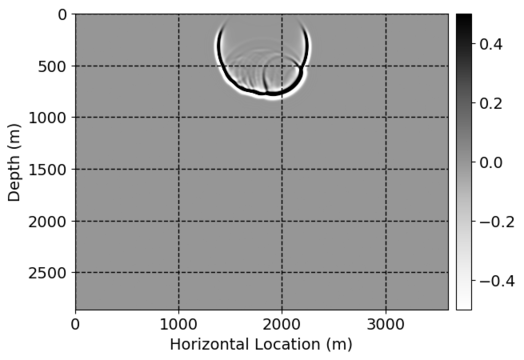}}
\subfloat[\label{corr-10-dsp-2}]{\includegraphics[width=0.250\hsize]{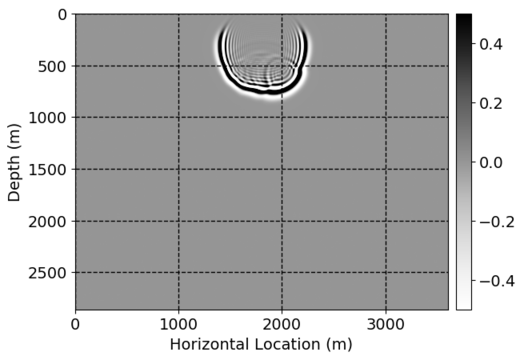}}
\subfloat[\label{corr-10-res-2}]{\includegraphics[width=0.250\hsize]{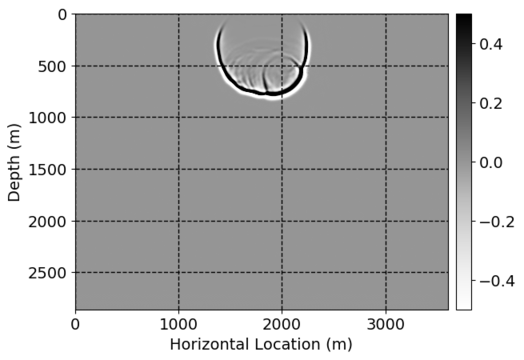}}
\subfloat[\label{corr-10-err-2}]{\includegraphics[width=0.250\hsize]{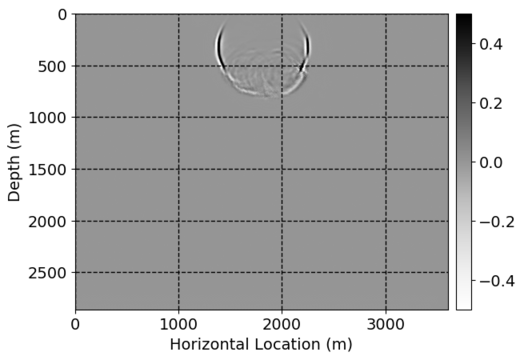}}
\\
\subfloat[\label{corr-10-true-3}]{\includegraphics[width=0.250\hsize]{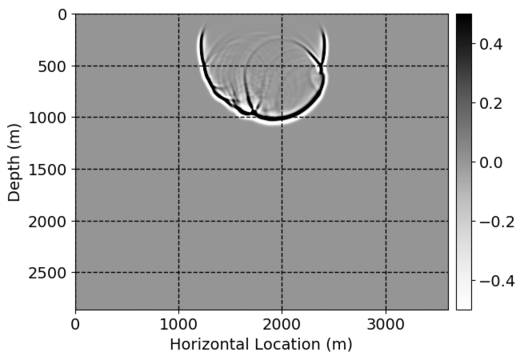}}
\subfloat[\label{corr-10-dsp-3}]{\includegraphics[width=0.250\hsize]{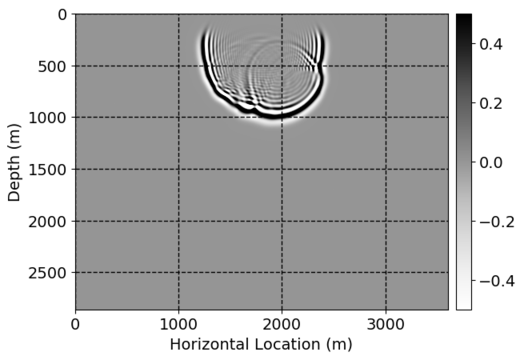}}
\subfloat[\label{corr-10-res-3}]{\includegraphics[width=0.250\hsize]{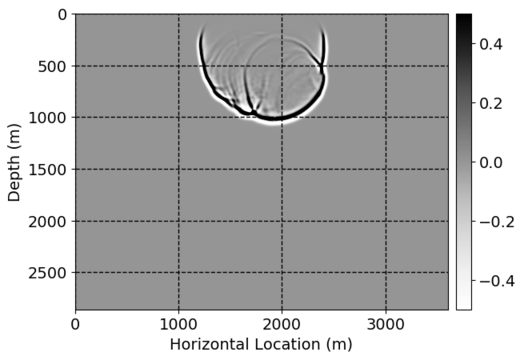}}
\subfloat[\label{corr-10-err-3}]{\includegraphics[width=0.250\hsize]{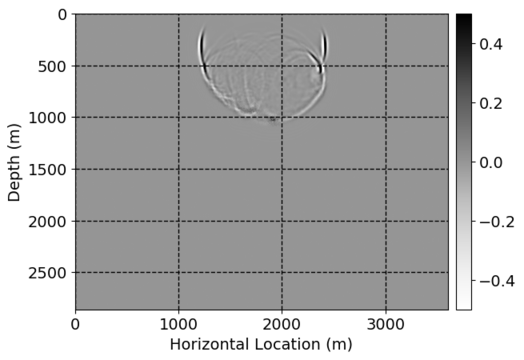}}
\\
\subfloat[\label{corr-10-true-4}]{\includegraphics[width=0.250\hsize]{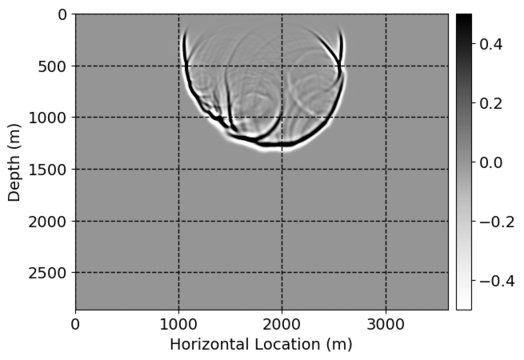}}
\subfloat[\label{corr-10-dsp-4}]{\includegraphics[width=0.250\hsize]{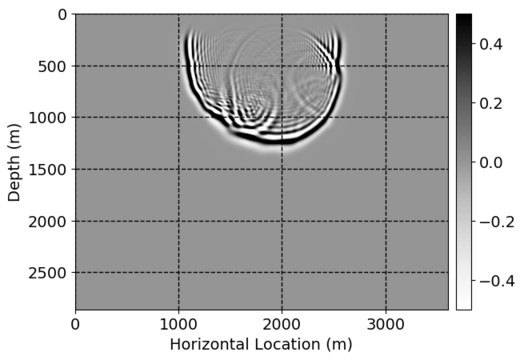}}
\subfloat[\label{corr-10-res-4}]{\includegraphics[width=0.250\hsize]{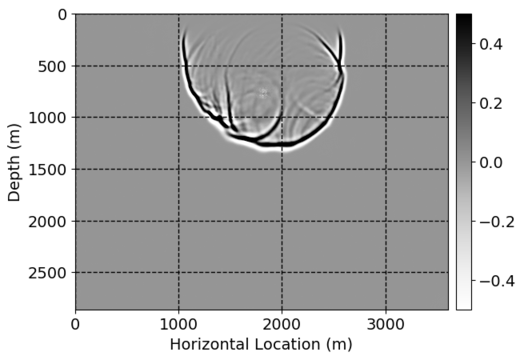}}
\subfloat[\label{corr-10-err-4}]{\includegraphics[width=0.250\hsize]{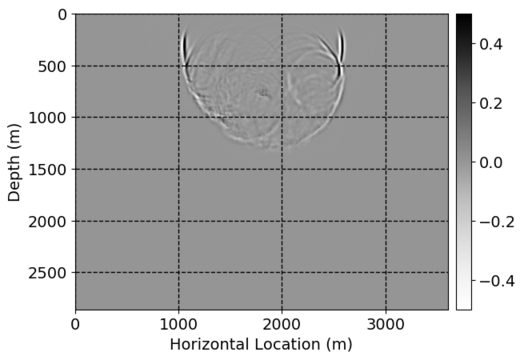}}
\caption{Neural network augmented wave simulation with ten CNNs, first
part. First column from top to bottom: a) to q) high-fidelity wavefield
snapshots, in order. Second column from top to bottom: b) to r)
low-fidelity wavefield snapshots simulated by solving
Equation~\ref{low-fidelity} with the same simulation time as
high-fidelity wavefields, in order. Third column from top to bottom: c)
to s) result of neural network augmented wave-equation simulation.
Output of the first to the fifth CNN, in order. Fourth column from top
to bottom: d) to t) difference between first and third column, in
order.}\label{corr-10-1}
\end{figure}

\begin{figure}
\centering
\subfloat[\label{corr-10-true-5}]{\includegraphics[width=0.250\hsize]{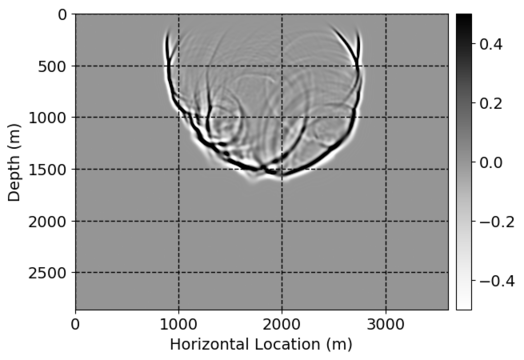}}
\subfloat[\label{corr-10-dsp-5}]{\includegraphics[width=0.250\hsize]{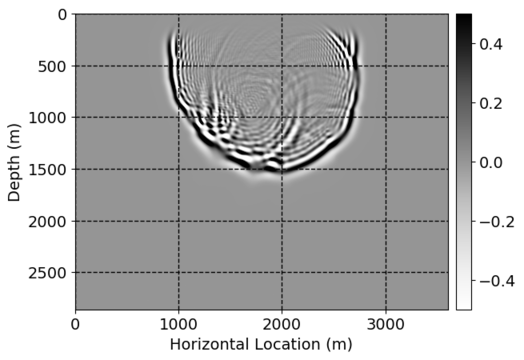}}
\subfloat[\label{corr-10-res-5}]{\includegraphics[width=0.250\hsize]{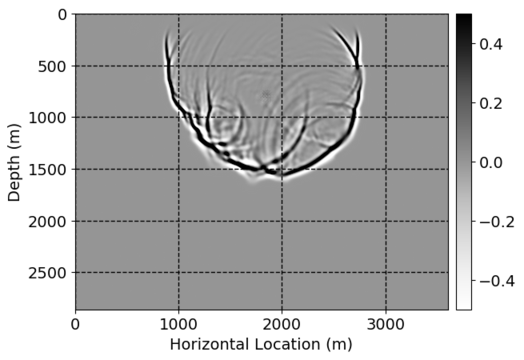}}
\subfloat[\label{corr-10-err-5}]{\includegraphics[width=0.250\hsize]{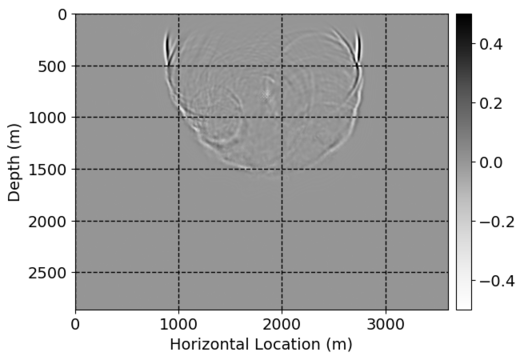}}
\\
\subfloat[\label{corr-10-true-6}]{\includegraphics[width=0.250\hsize]{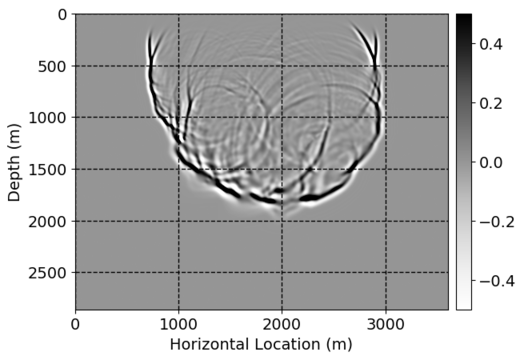}}
\subfloat[\label{corr-10-dsp-6}]{\includegraphics[width=0.250\hsize]{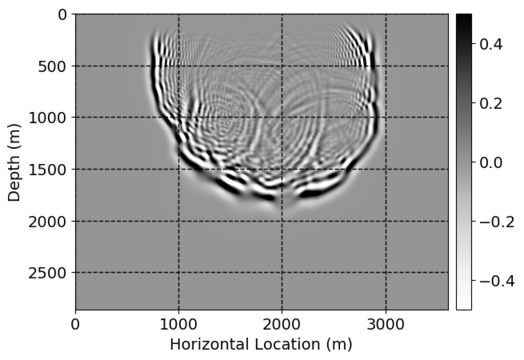}}
\subfloat[\label{corr-10-res-6}]{\includegraphics[width=0.250\hsize]{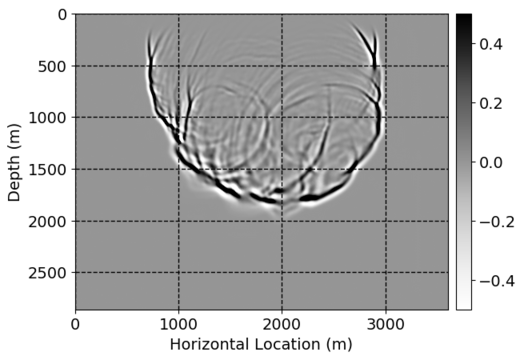}}
\subfloat[\label{corr-10-err-6}]{\includegraphics[width=0.250\hsize]{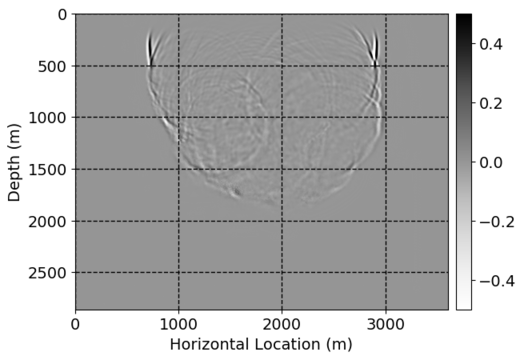}}
\\
\subfloat[\label{corr-10-true-7}]{\includegraphics[width=0.250\hsize]{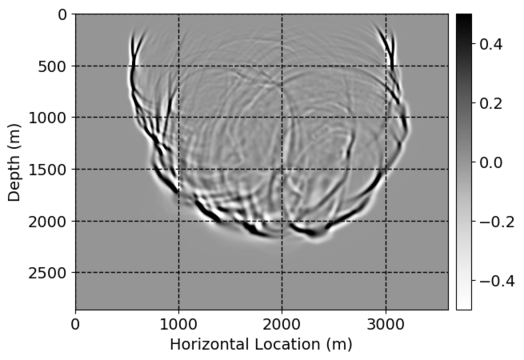}}
\subfloat[\label{corr-10-dsp-7}]{\includegraphics[width=0.250\hsize]{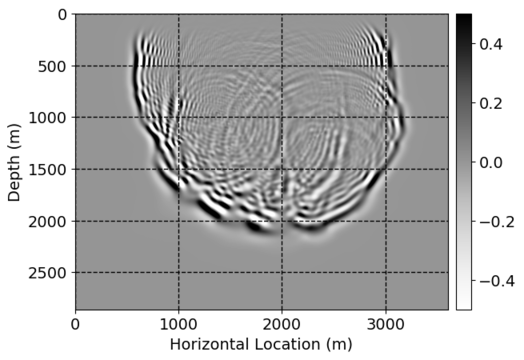}}
\subfloat[\label{corr-10-res-7}]{\includegraphics[width=0.250\hsize]{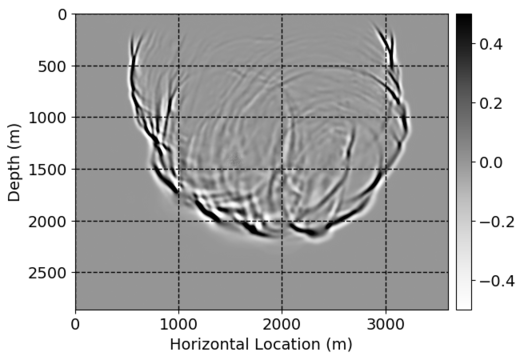}}
\subfloat[\label{corr-10-err-7}]{\includegraphics[width=0.250\hsize]{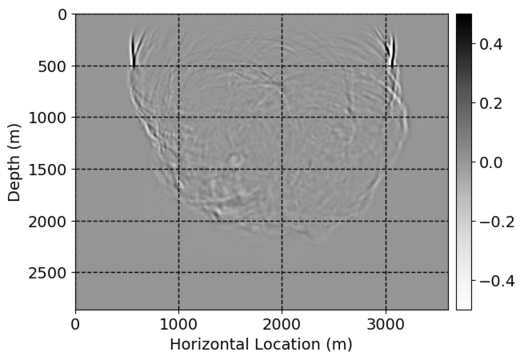}}
\\
\subfloat[\label{corr-10-true-8}]{\includegraphics[width=0.250\hsize]{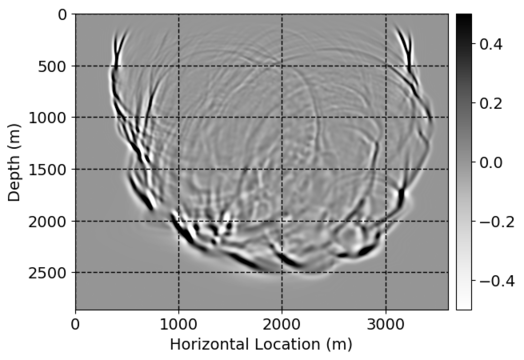}}
\subfloat[\label{corr-10-dsp-8}]{\includegraphics[width=0.250\hsize]{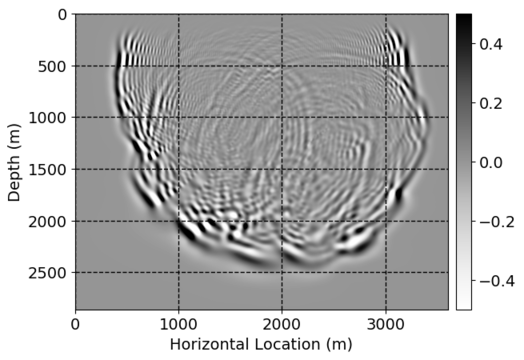}}
\subfloat[\label{corr-10-res-8}]{\includegraphics[width=0.250\hsize]{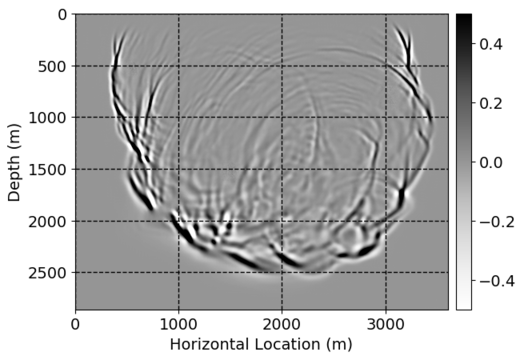}}
\subfloat[\label{corr-10-err-8}]{\includegraphics[width=0.250\hsize]{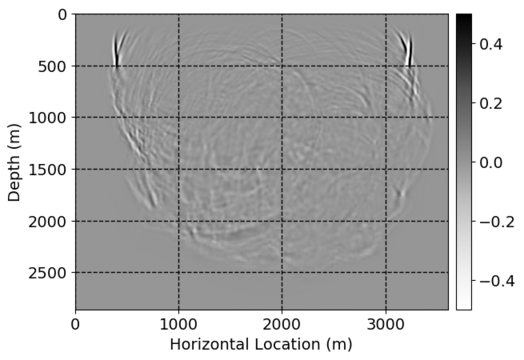}}
\\
\subfloat[\label{corr-10-true-9}]{\includegraphics[width=0.250\hsize]{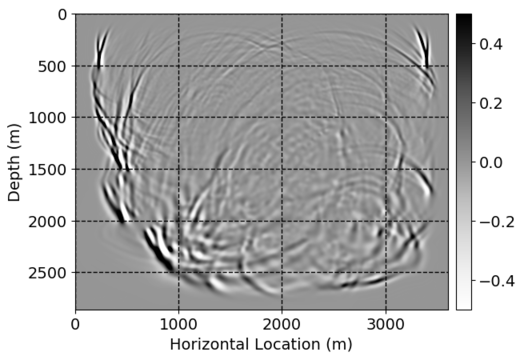}}
\subfloat[\label{corr-10-dsp-9}]{\includegraphics[width=0.250\hsize]{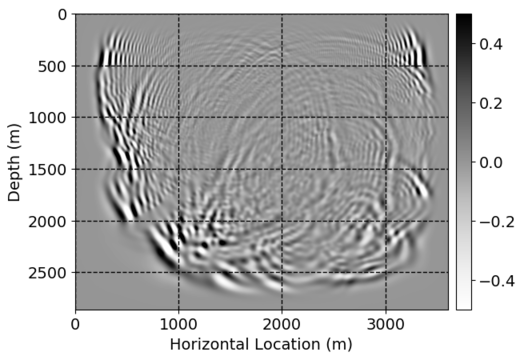}}
\subfloat[\label{corr-10-res-9}]{\includegraphics[width=0.250\hsize]{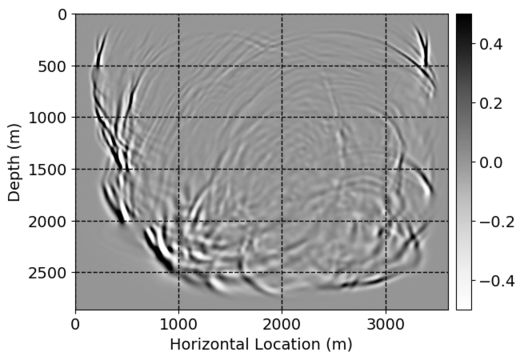}}
\subfloat[\label{corr-10-err-9}]{\includegraphics[width=0.250\hsize]{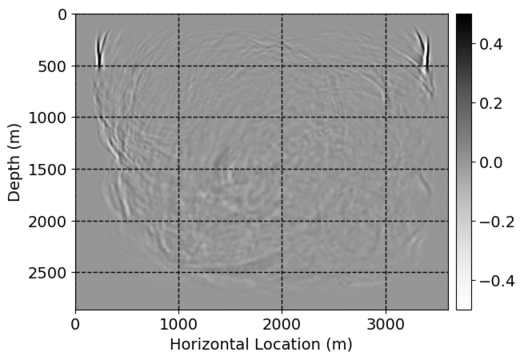}}
\caption{Neural network augmented wave simulation with ten CNNs, second
part. First column from top to bottom: a) to q) high-fidelity wavefield
snapshots, in order. Second column from top to bottom: b) to r)
low-fidelity wavefield snapshots simulated by solving
Equation~\ref{low-fidelity} with the same simulation time as
high-fidelity wavefields, in order. Third column from top to bottom: c)
to s) result of neural network augmented wave-equation simulation.
Output of the sixth to the last CNN, in order. Fourth column from top to
bottom: d) to t) difference between first and third column, in
order.}\label{corr-10-2}
\end{figure}

Next, we will demonstrate the the corrected wavefields in three
conducted experiments evaluated over one testing shot location. For each
experiment, we show the high-fidelity wavefield snapshots,
$\mathbf{u}_{\tau_i}, \ i = 0, 1, \ldots , \lfloor \frac{M-1}{k} \rfloor$,
where $i$ iterates over the CNNs, numerically dispersed low-fidelity
wavefields, and the corrected wavefield snapshots by the CNNs. To
evaluate the performance of each correction, we also depict the
correction error---i.e., difference between the high-fidelity and
corrected wavefield snapshots. Figure~\ref{corr-3} shows the mentioned
wavefield snapshots for the neural network augmented wave simulator with
three CNNs. First column shows the high-fidelity wavefields by solving
Equation~\ref{high-fidelity}, second column depicts low-fidelity
simulations by solving Equation~\ref{low-fidelity}, third column
indicates the result of neural network augmented wavefield simulations,
and the fourth column is the learned wave simulation error---i.e.,
difference between the first and last column in Figure~\ref{corr-3}.
Similarly, Figures~\ref{corr-5} $-$ \ref{corr-10-2} show the high- and
low- fidelity and learned wave simulation wavefield snapshots in the
first three columns, in order, for the neural network augmented wave
simulator with five and ten CNNs, respectively.

As expected because of the reasons stated before, we observe that the
quality of neural network augmented wave-equation simulation degrades as
the number of CNNs increases. On the other hands, the high quality of
learned wave simulation with few CNNs (see Figure~\ref{corr-3}) suggests
the quality of the simulation with more CNNs might be improved by
increasing the number of iterations. As it can be seen in the last
column of Figures~\ref{corr-10-1} and~\ref{corr-10-2}, the learned wave
simulation with ten CNNs has the lowest quality. It can be seen that the
learned wave simulation has the least accuracy in direct wave, which
happen to be the events with largest amplitudes. Also, it appears that
most of the numerical dispersion has been removed, the phase has been
recovered, and residual is mostly amplitude differences.

\subsection{Performance comparison: Single CNN low-to-high-fidelity
mapping}\label{performance-comparison-single-cnn-low-to-high-fidelity-mapping}

In order to evaluate the effectiveness of the proposed method we also
train a single CNN similar to our previous attempt to remove numerical
dispersion from wavefield snapshots
\citep{siahkoohi2018deep, siahkoohi2019transfer} and compare the result
of numerical dispersion removal with the proposed method. To be more
precise, for each presented neural network augmented wave-equation
simulation experiment, where we use three, five, and ten CNNs, we train
a single CNN, $\mathcal{G}_{\theta}$, with the same architecture as the
architecture used in learned wave propagators, in order to remove
numerical dispersion from all the low-fidelity wavefield snapshots
simulated by solving Equation~\ref{low-fidelity} for
$j\equiv k-1 \ (\bmod \ k)$, on training shot locations. Likewise to
previous examples, here we also use half of the available shot locations
to simulate training pairs, and the rest is used to evaluate the
performance of the trained CNN. The input to $\mathcal{G}_{\theta}$
during training can be written as follows (compare with
Equation~\ref{inputtoCNN}):
\begin{equation}
\begin{aligned}
    &\ \mathbf{\tilde{u}}_i = \bar{F}_k (\mathbf{\tilde{u}}_{i-1} ), \quad i = 1, 2, \ldots , \lfloor \frac{M-1}{k} \rfloor, \\
    &\ \mathbf{\tilde{u}}_0 = \bar{F}_k(\mathbf{q}), \\
\end{aligned}
\label{inputtoSingleCNN}
\end{equation}
 The desired output for the mentioned CNN is the high-fidelity wavefield
snapshots simulated on training shot locations,
$\mathbf{u}_{\tau_i}^{(p)}, \ p = 0,1, \ldots , n-1, \ i = 0, 1, \ldots , \lfloor \frac{M-1}{k} \rfloor$.
The objective function for the mentioned CNN can be represented as
follows:
\begin{equation}
\begin{split}
    \mathcal{L} = \frac{1}{n (\lfloor \frac{M-1}{k} \rfloor + 1)}  \sum_{i=0}^{\lfloor \frac{M-1}{k} \rfloor} \sum_{p=0}^{n-1} \left \| \mathcal{G}_{\theta}(\mathbf{\tilde{u}}_i^{(p)}) - \mathbf{u}_{\tau_i}^{(p)} \right \|_1. \\
\end{split}
\label{trainCNN-single}
\end{equation}
 We minimize objective function~\ref{trainCNN-single} over $\theta$ with
Adam optimizes, using the same maximum number of iterations as before,
this time by combining all the training pairs associated with different
CNNs in the learned wavefield simulation example. As mentioned before,
in order to compare with the proposed method, we minimize objective
function~\ref{trainCNN-single} over three different set of input-output
pairs, each corresponding to our presented experiments with varying
number of CNNs. Table~\ref{training-details-compare} summarizes the
total number of iterations, training pairs, training time, and number of
tunable parameters for three different cases, which differ in number of
timesteps which we choose to correct the numerical dispersion. This
selected timesteps are associated with the timesteps that the CNNs
operated on, in our three previous examples.

\begin{table}
\centering
\begin{tabular}{cccccc}
\toprule\addlinespace
\# Timesteps to correct & Iterations & Pairs per CNN & Time & Param.
count\tabularnewline
\midrule
$3$ & $100500$ & $603$ & $13.85$ hours & $11383424$\tabularnewline
$5$ & $100500$ & $1005$ & $13.96$ hours & $11383424$\tabularnewline
$10$ & $100500$ & $2010$ & $14.56$ hours & $11383424$\tabularnewline
\bottomrule
\end{tabular}
\caption{Summary of details in three conducted
experiments}\label{training-details-compare}
\end{table}

\begin{figure}
\centering
\subfloat[\label{compare-3-snr}]{\includegraphics[width=0.500\hsize]{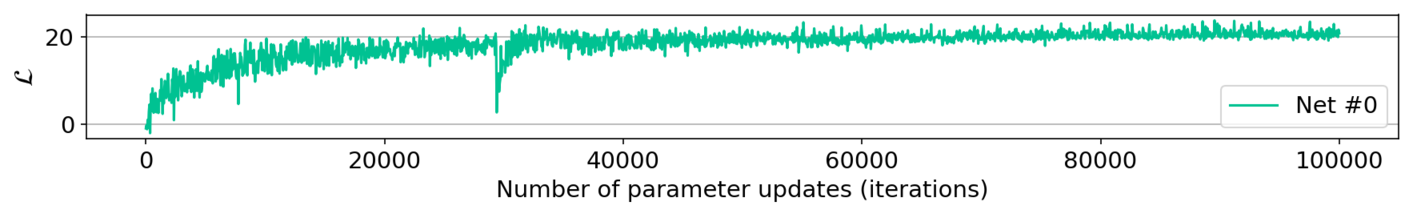}}
\subfloat[\label{compare-3-loss}]{\includegraphics[width=0.500\hsize]{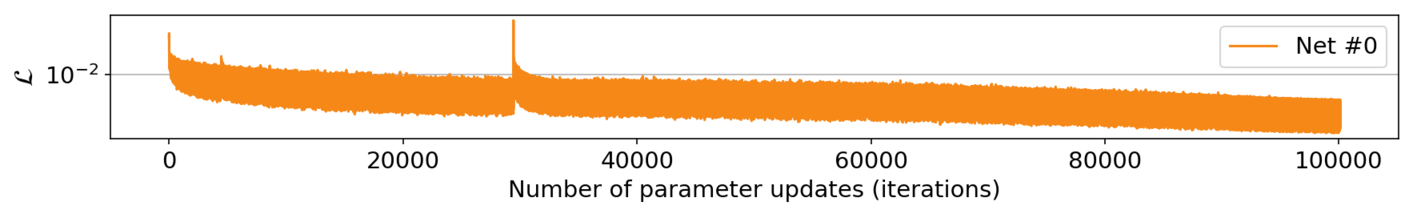}}
\\
\subfloat[\label{compare-5-snr}]{\includegraphics[width=0.500\hsize]{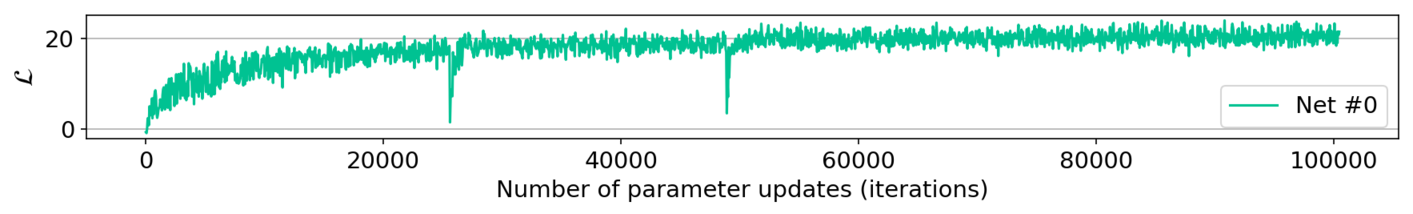}}
\subfloat[\label{compare-5-loss}]{\includegraphics[width=0.500\hsize]{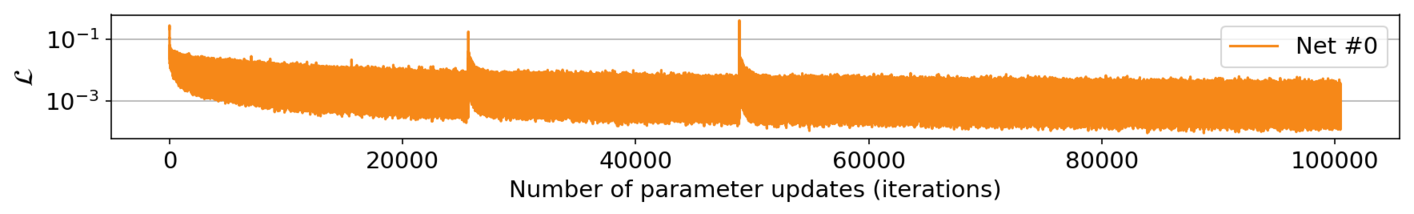}}
\\
\subfloat[\label{compare-10-snr}]{\includegraphics[width=0.500\hsize]{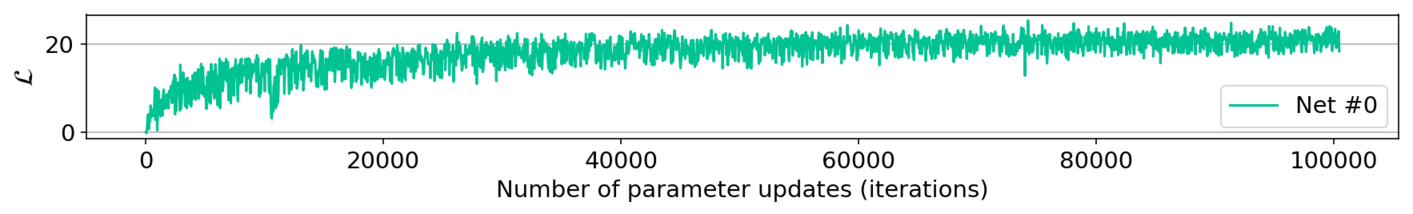}}
\subfloat[\label{compare-10-loss}]{\includegraphics[width=0.500\hsize]{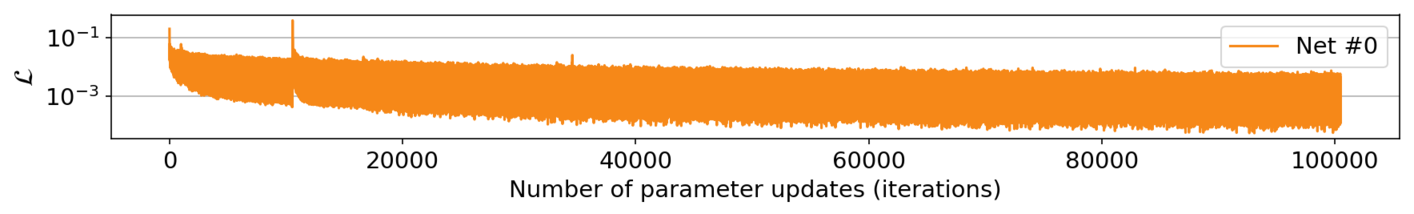}}
\caption{Single CNN low-to-high-fidelity mapping. a) wavefield snapshot
correction SNR curve and b) objective function value curve when CNN is
trained on wavefield snapshot pairs corresponding to learned wavefield
simulation with three CNNs. c) wavefield snapshot correction SNR curve
and d) objective function value curve when CNN is trained on wavefield
snapshot pairs corresponding to learned wavefield simulation with five
CNNs. e) wavefield snapshot correction SNR curve and f) objective
function value curve when CNN is trained on wavefield snapshot pairs
corresponding to learned wavefield simulation with ten
CNNs.}\label{compare-log}
\end{figure}

The slight difference in runtime among three different cases provided in
Table~\ref{training-details-compare} is partly due to different number
of training pairs needed to be generated. Figure~\ref{compare-log}
depicts the wavefield snapshot correction SNR curves, evaluated on
testing pairs while training, and the value of objective
function~\ref{trainCNN-single}, in single CNN low-to-high-fidelity
mapping experiment, as a function of number of iterations.
Figures~\ref{compare-3-snr}, \ref{compare-5-snr},
and~\ref{compare-10-snr} show the SNR curves, when we trained the CNN on
wavefield snapshots correspond to three, five, and ten, timesteps,
respectively. Similarly, Figures~\ref{compare-3-loss},
\ref{compare-5-loss}, and~\ref{compare-10-loss} depict the training
objective function value (Equation~\ref{trainCNN-single}), when the CNN
is trained on wavefield snapshots correspond to three, five, and ten,
timesteps, respectively. SNR curves depicted in first column of
Figure~\ref{compare-log} show the evolution of wavefield correction SNR
evaluated on randomly selected testing wavefield wavefields form the
wavefield snapshots combined from different timesteps. Therefore,
Figures~\ref{compare-3-snr},
\ref{compare-5-snr},~and~\ref{compare-10-snr} indicate that the three
different CNNs converge to a wavefield correction SNR around $20$ dB,
regardless of number of timesteps they are correcting for. Although this
does not suggest that the performance will stay the same as we increase
the number of timesteps needed to be corrected. By comparing
Figure~\ref{compare-3-snr} with first column of Figure~\ref{corr-3-log}
(SNR curves for neural network augmented wave-equation simulation with
three CNNs), we observe that, on average the two methods are performing
equally well.

\begin{figure}
\centering
\subfloat[\label{compare-3-true-0}]{\includegraphics[width=0.250\hsize]{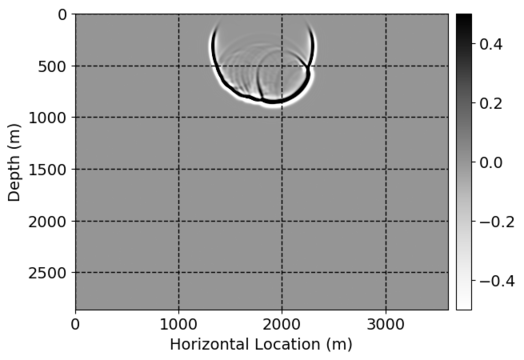}}
\subfloat[\label{compare-3-dsp-0}]{\includegraphics[width=0.250\hsize]{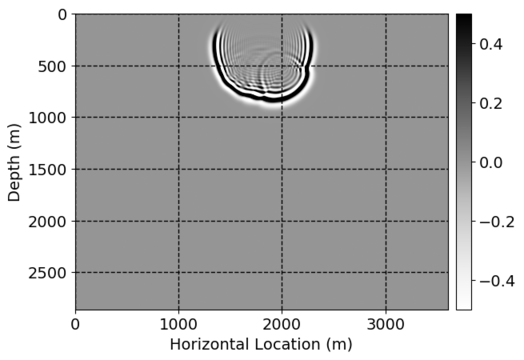}}
\subfloat[\label{compare-3-res-0}]{\includegraphics[width=0.250\hsize]{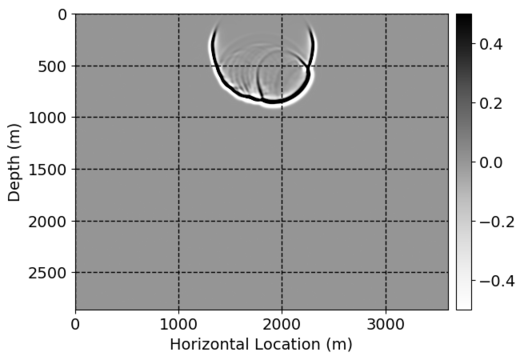}}
\subfloat[\label{compare-3-err-0}]{\includegraphics[width=0.250\hsize]{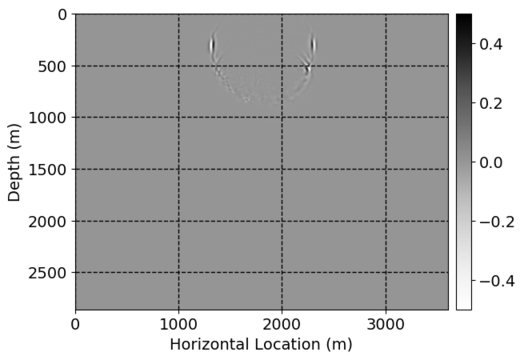}}
\\
\subfloat[\label{compare-3-true-1}]{\includegraphics[width=0.250\hsize]{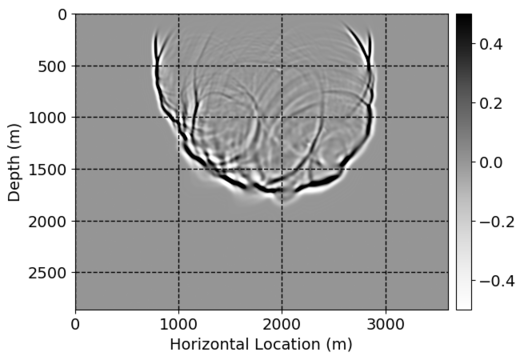}}
\subfloat[\label{compare-3-dsp-1}]{\includegraphics[width=0.250\hsize]{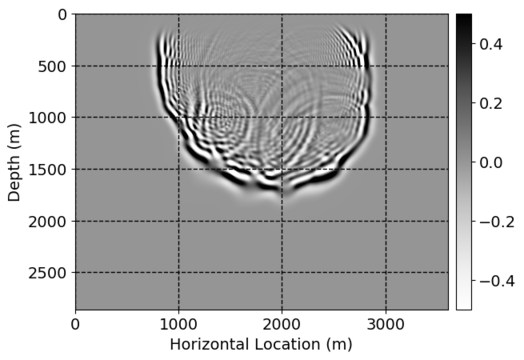}}
\subfloat[\label{compare-3-res-1}]{\includegraphics[width=0.250\hsize]{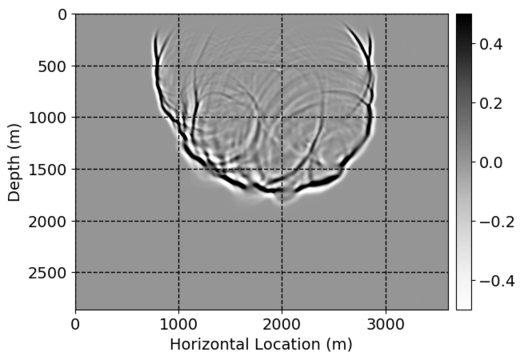}}
\subfloat[\label{compare-3-err-1}]{\includegraphics[width=0.250\hsize]{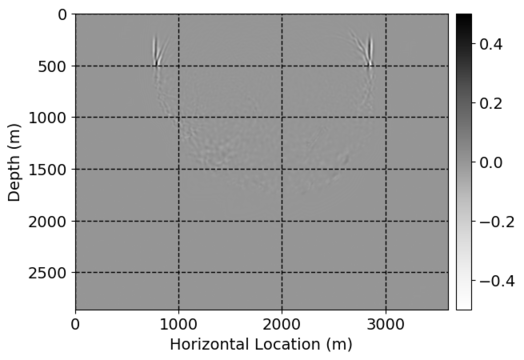}}
\\
\subfloat[\label{compare-3-true-2}]{\includegraphics[width=0.250\hsize]{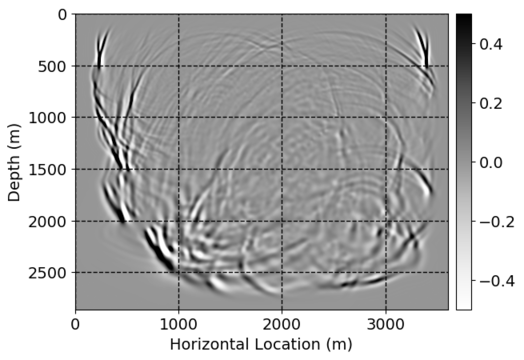}}
\subfloat[\label{compare-3-dsp-2}]{\includegraphics[width=0.250\hsize]{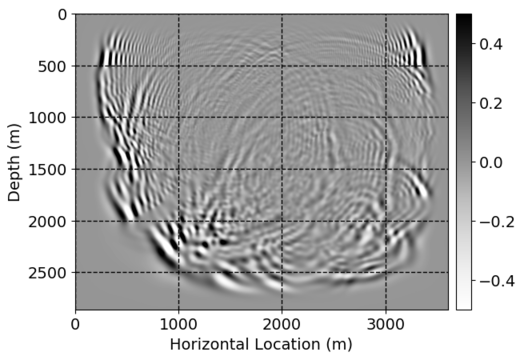}}
\subfloat[\label{compare-3-res-2}]{\includegraphics[width=0.250\hsize]{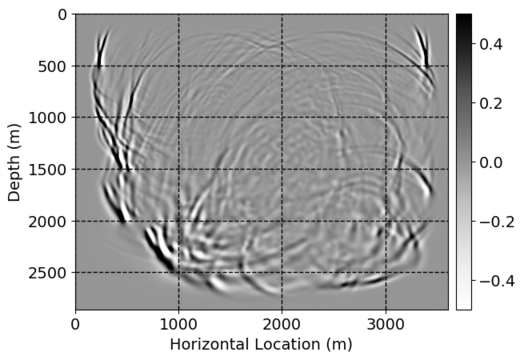}}
\subfloat[\label{compare-3-err-2}]{\includegraphics[width=0.250\hsize]{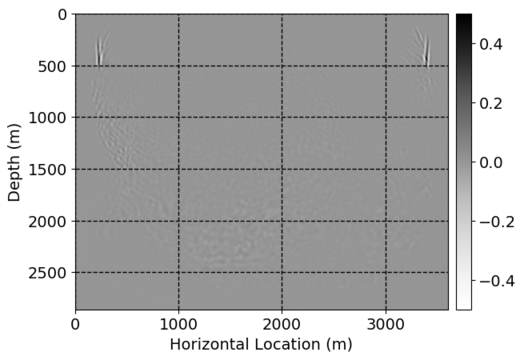}}
\caption{Single CNN low-to-high-fidelity mapping with three timesteps to
be corrected. First column from top to bottom: a, e, i) high-fidelity
wavefield snapshots, in order. Second column from top to bottom: b, f,
j) low-fidelity wavefield snapshots simulated by solving
Equation~\ref{low-fidelity} with the same simulation time as
high-fidelity wavefields, in order. Third column from top to bottom: c,
g, k) result of single CNN low-to-high-fidelity mapping. Fourth column
from top to bottom: d, h, l) difference between first and third column,
in order.}\label{compare-3}
\end{figure}

\begin{figure}
\centering
\subfloat[\label{compare-5-true-0}]{\includegraphics[width=0.250\hsize]{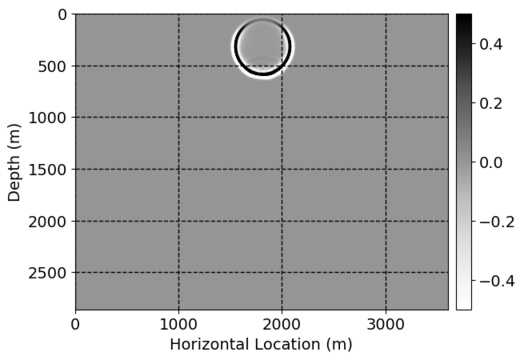}}
\subfloat[\label{compare-5-dsp-0}]{\includegraphics[width=0.250\hsize]{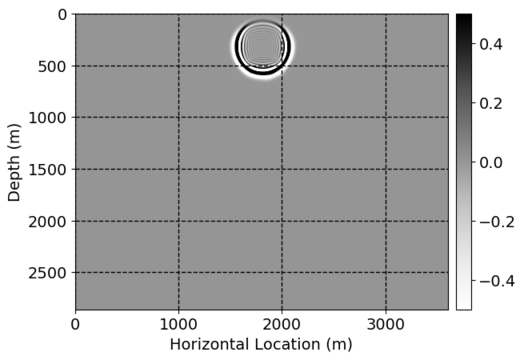}}
\subfloat[\label{compare-5-res-0}]{\includegraphics[width=0.250\hsize]{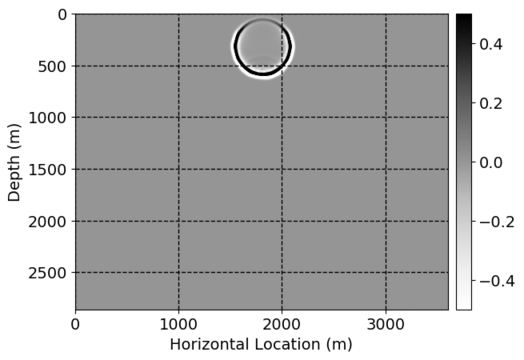}}
\subfloat[\label{compare-5-err-0}]{\includegraphics[width=0.250\hsize]{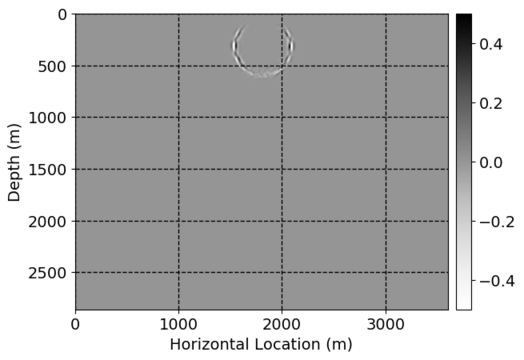}}
\\
\subfloat[\label{compare-5-true-1}]{\includegraphics[width=0.250\hsize]{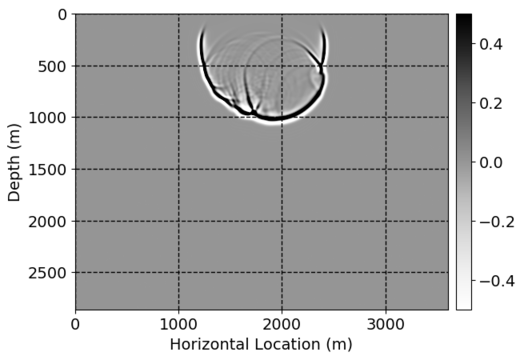}}
\subfloat[\label{compare-5-dsp-1}]{\includegraphics[width=0.250\hsize]{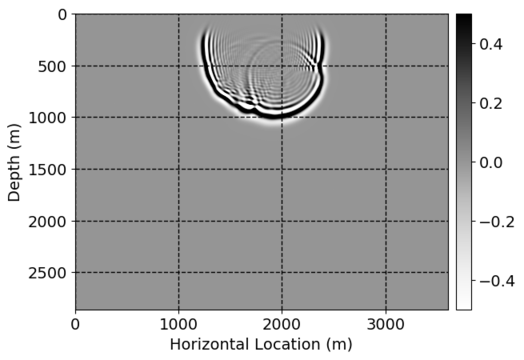}}
\subfloat[\label{compare-5-res-1}]{\includegraphics[width=0.250\hsize]{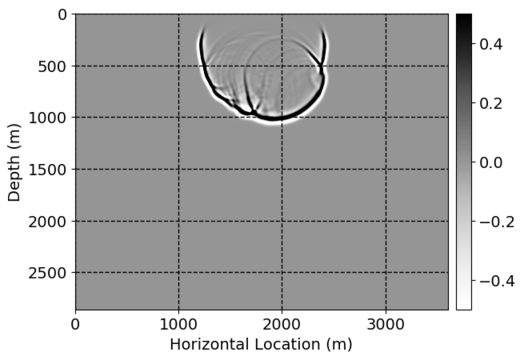}}
\subfloat[\label{compare-5-err-1}]{\includegraphics[width=0.250\hsize]{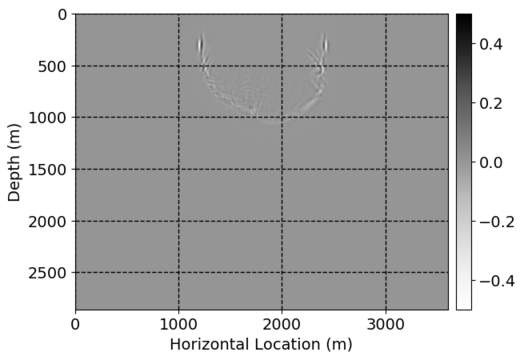}}
\\
\subfloat[\label{compare-5-true-2}]{\includegraphics[width=0.250\hsize]{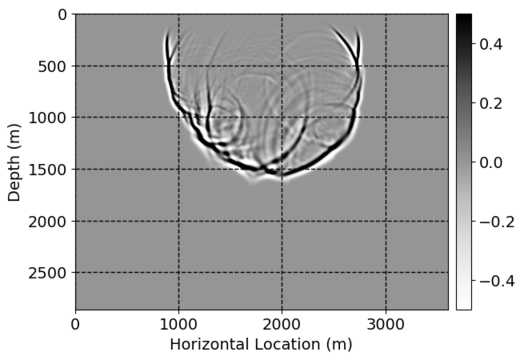}}
\subfloat[\label{compare-5-dsp-2}]{\includegraphics[width=0.250\hsize]{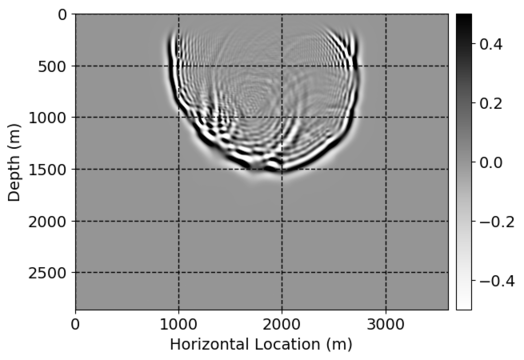}}
\subfloat[\label{compare-5-res-2}]{\includegraphics[width=0.250\hsize]{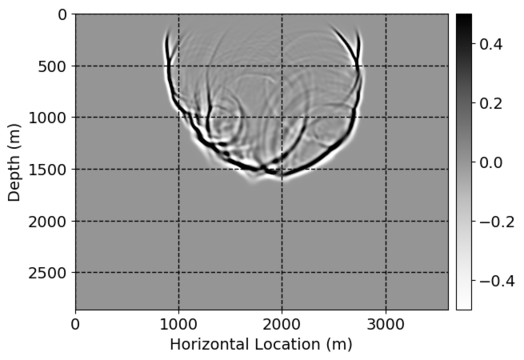}}
\subfloat[\label{compare-5-err-2}]{\includegraphics[width=0.250\hsize]{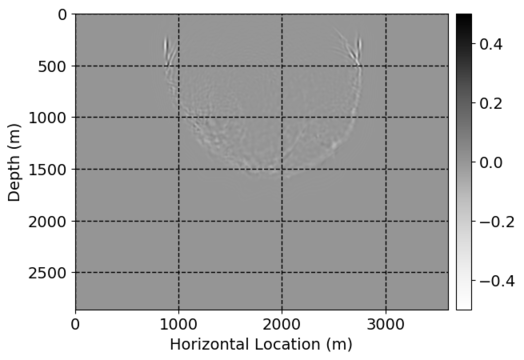}}
\\
\subfloat[\label{compare-5-true-3}]{\includegraphics[width=0.250\hsize]{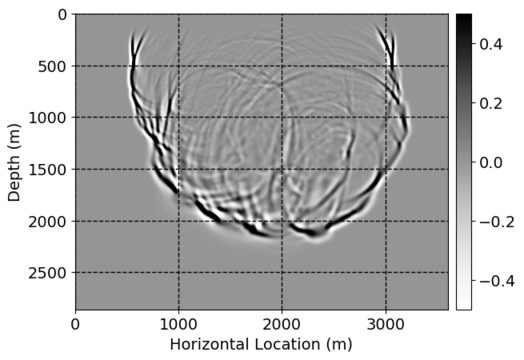}}
\subfloat[\label{compare-5-dsp-3}]{\includegraphics[width=0.250\hsize]{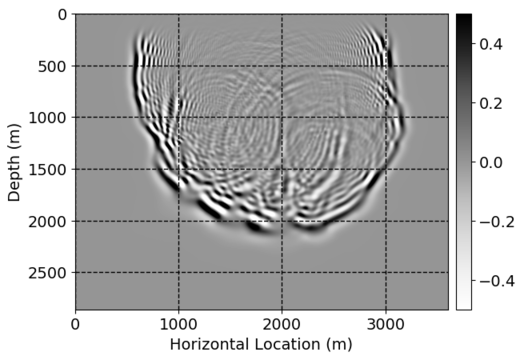}}
\subfloat[\label{compare-5-res-3}]{\includegraphics[width=0.250\hsize]{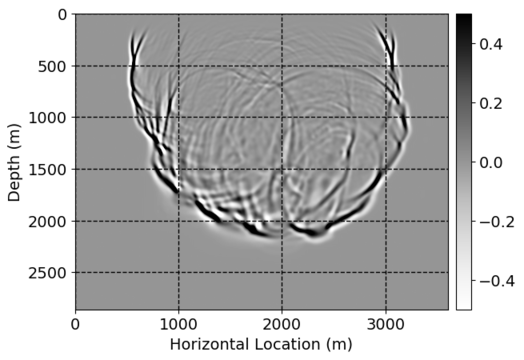}}
\subfloat[\label{compare-5-err-3}]{\includegraphics[width=0.250\hsize]{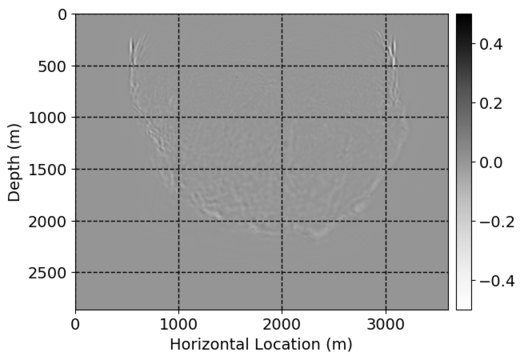}}
\\
\subfloat[\label{compare-5-true-4}]{\includegraphics[width=0.250\hsize]{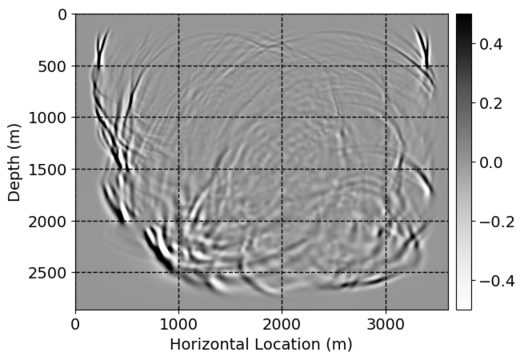}}
\subfloat[\label{compare-5-dsp-4}]{\includegraphics[width=0.250\hsize]{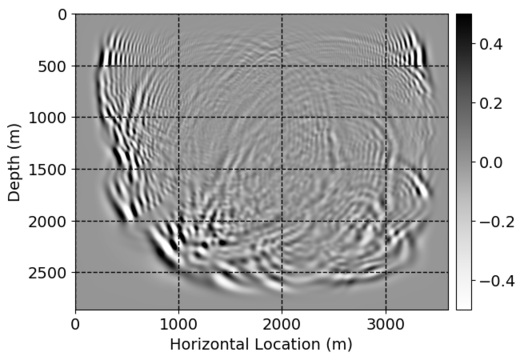}}
\subfloat[\label{compare-5-res-4}]{\includegraphics[width=0.250\hsize]{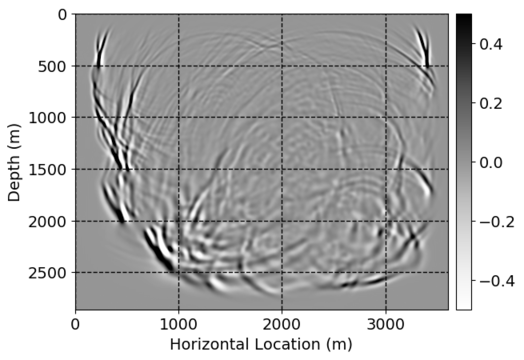}}
\subfloat[\label{compare-5-err-4}]{\includegraphics[width=0.250\hsize]{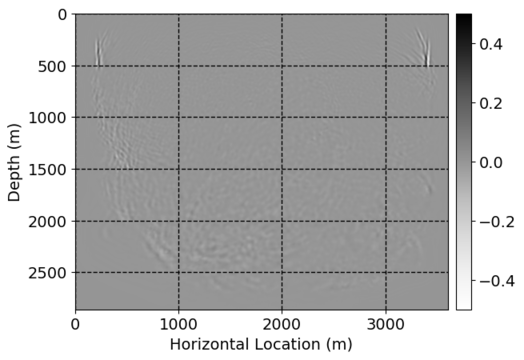}}
\caption{Single CNN low-to-high-fidelity mapping with five timesteps to
be corrected. First column from top to bottom: a) to q) high-fidelity
wavefield snapshots, in order. Second column from top to bottom: b) to
r) low-fidelity wavefield snapshots simulated by solving
Equation~\ref{low-fidelity} with the same simulation time as
high-fidelity wavefields, in order. Third column from top to bottom: c)
to s) result of single CNN low-to-high-fidelity mapping. Fourth column
from top to bottom: d) to t) difference between first and third column,
in order.}\label{compare-5}
\end{figure}

\begin{figure}
\centering
\subfloat[\label{compare-10-true-0}]{\includegraphics[width=0.250\hsize]{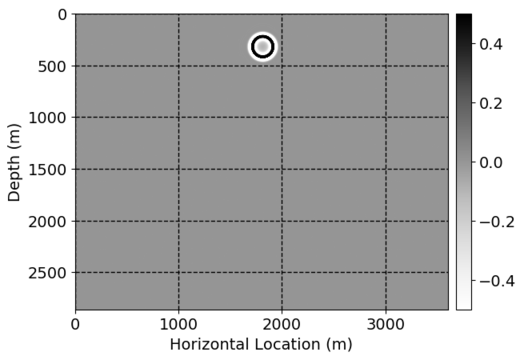}}
\subfloat[\label{compare-10-dsp-0}]{\includegraphics[width=0.250\hsize]{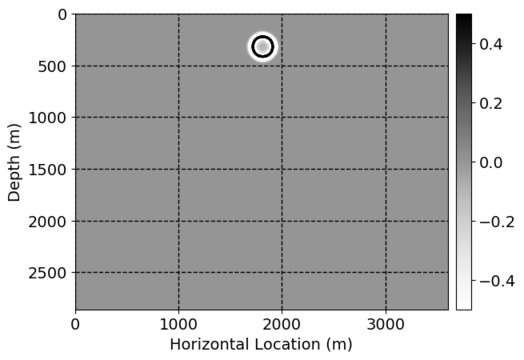}}
\subfloat[\label{compare-10-res-0}]{\includegraphics[width=0.250\hsize]{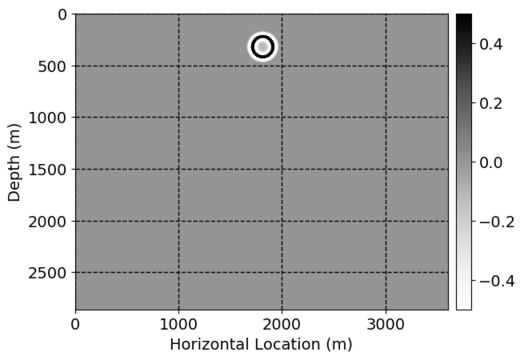}}
\subfloat[\label{compare-10-err-0}]{\includegraphics[width=0.250\hsize]{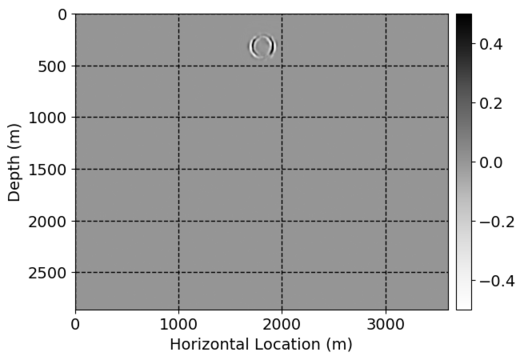}}
\\
\subfloat[\label{compare-10-true-1}]{\includegraphics[width=0.250\hsize]{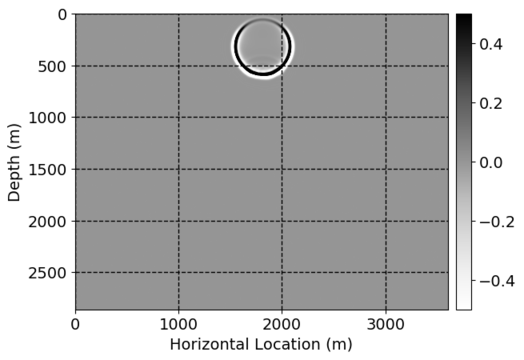}}
\subfloat[\label{compare-10-dsp-1}]{\includegraphics[width=0.250\hsize]{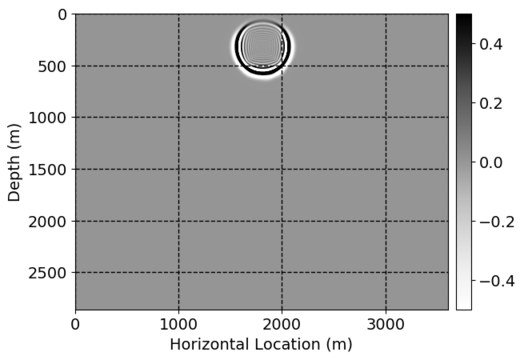}}
\subfloat[\label{compare-10-res-1}]{\includegraphics[width=0.250\hsize]{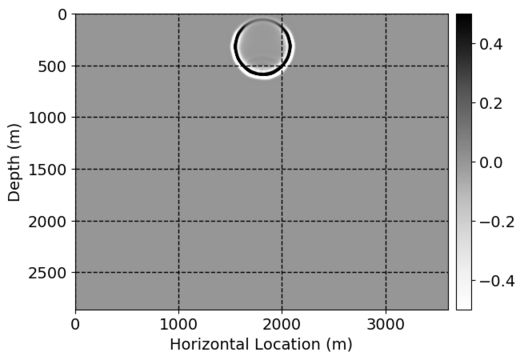}}
\subfloat[\label{compare-10-err-1}]{\includegraphics[width=0.250\hsize]{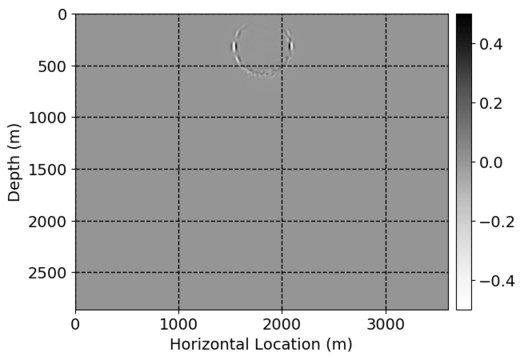}}
\\
\subfloat[\label{compare-10-true-2}]{\includegraphics[width=0.250\hsize]{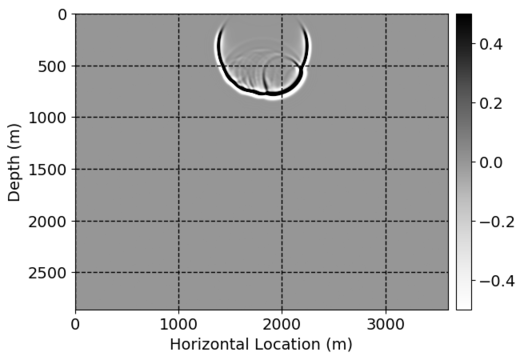}}
\subfloat[\label{compare-10-dsp-2}]{\includegraphics[width=0.250\hsize]{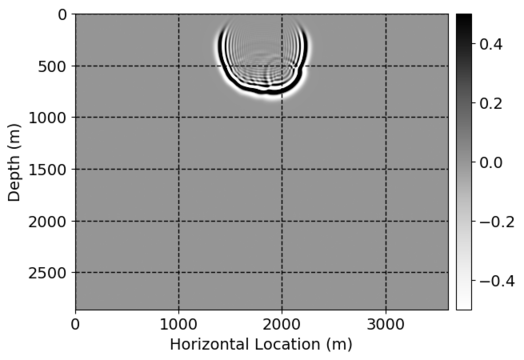}}
\subfloat[\label{compare-10-res-2}]{\includegraphics[width=0.250\hsize]{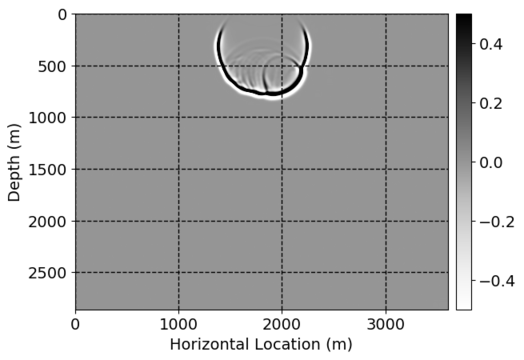}}
\subfloat[\label{compare-10-err-2}]{\includegraphics[width=0.250\hsize]{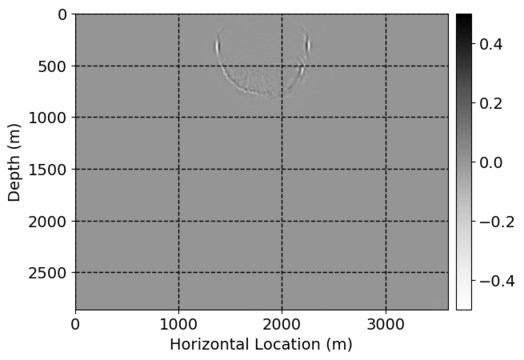}}
\\
\subfloat[\label{compare-10-true-3}]{\includegraphics[width=0.250\hsize]{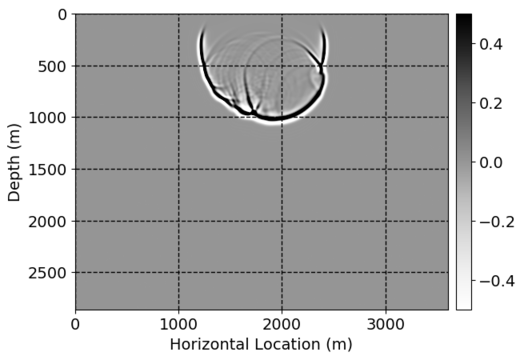}}
\subfloat[\label{compare-10-dsp-3}]{\includegraphics[width=0.250\hsize]{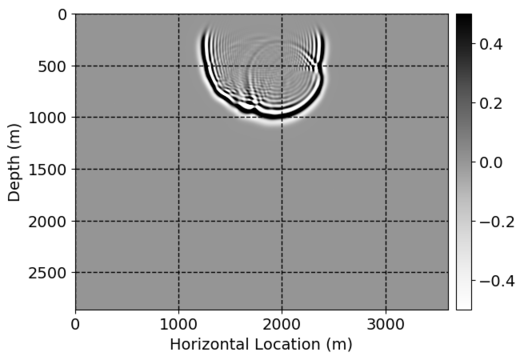}}
\subfloat[\label{compare-10-res-3}]{\includegraphics[width=0.250\hsize]{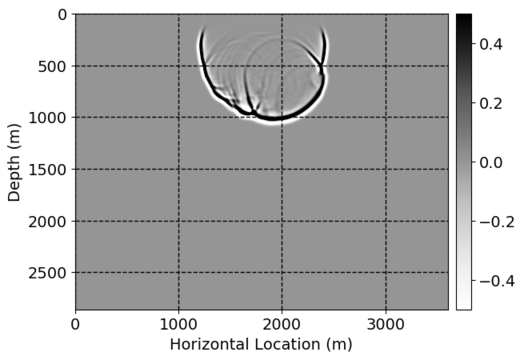}}
\subfloat[\label{compare-10-err-3}]{\includegraphics[width=0.250\hsize]{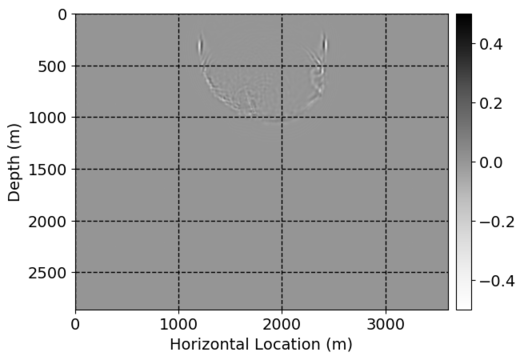}}
\\
\subfloat[\label{compare-10-true-4}]{\includegraphics[width=0.250\hsize]{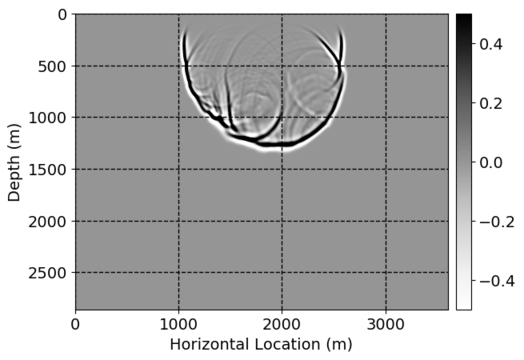}}
\subfloat[\label{compare-10-dsp-4}]{\includegraphics[width=0.250\hsize]{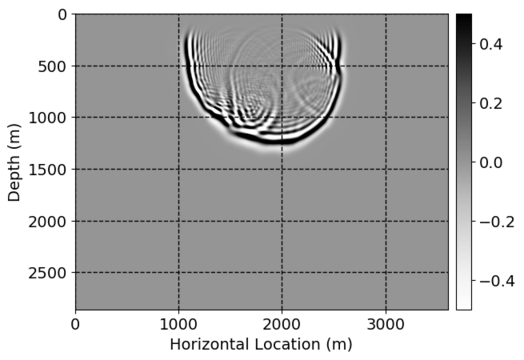}}
\subfloat[\label{compare-10-res-4}]{\includegraphics[width=0.250\hsize]{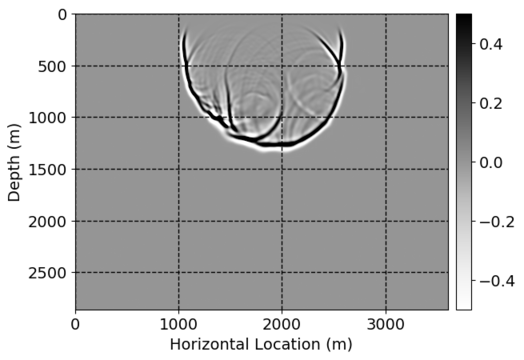}}
\subfloat[\label{compare-10-err-4}]{\includegraphics[width=0.250\hsize]{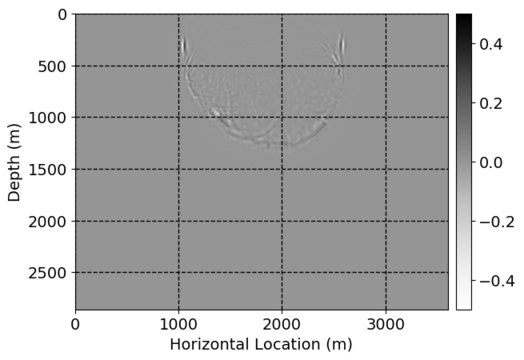}}
\caption{Single CNN low-to-high-fidelity mapping with ten timesteps to
be corrected, first part. First column from top to bottom: a) to q)
high-fidelity wavefield snapshots, in order. Second column from top to
bottom: b) to r) low-fidelity wavefield snapshots simulated by solving
Equation~\ref{low-fidelity} with the same simulation time as
high-fidelity wavefields, in order. Third column from top to bottom: c)
to s) result of single CNN low-to-high-fidelity mapping. Fourth column
from top to bottom: d) to t) difference between first and third column,
in order.}\label{compare-10-1}
\end{figure}

\begin{figure}
\centering
\subfloat[\label{compare-10-true-5}]{\includegraphics[width=0.250\hsize]{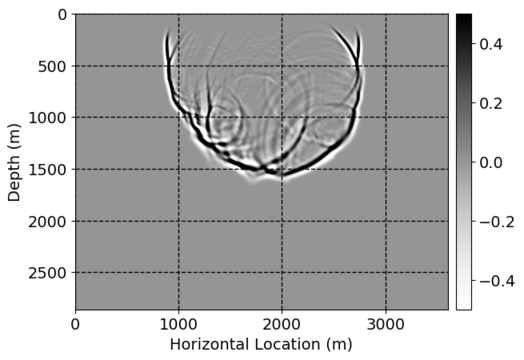}}
\subfloat[\label{compare-10-dsp-5}]{\includegraphics[width=0.250\hsize]{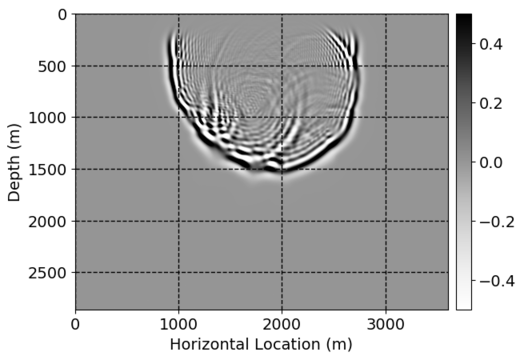}}
\subfloat[\label{compare-10-res-5}]{\includegraphics[width=0.250\hsize]{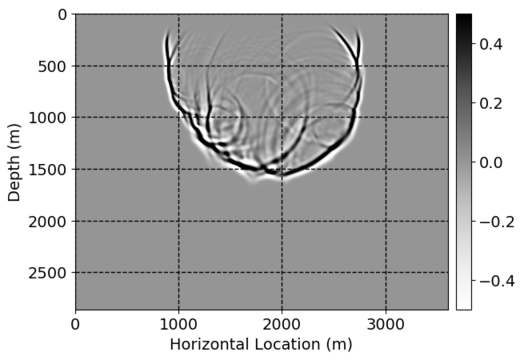}}
\subfloat[\label{compare-10-err-5}]{\includegraphics[width=0.250\hsize]{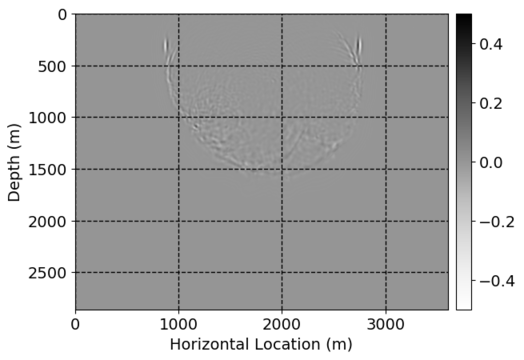}}
\\
\subfloat[\label{compare-10-true-6}]{\includegraphics[width=0.250\hsize]{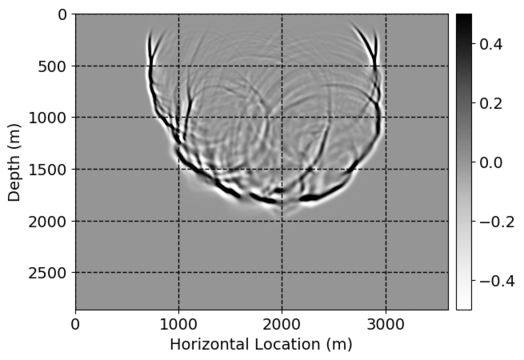}}
\subfloat[\label{compare-10-dsp-6}]{\includegraphics[width=0.250\hsize]{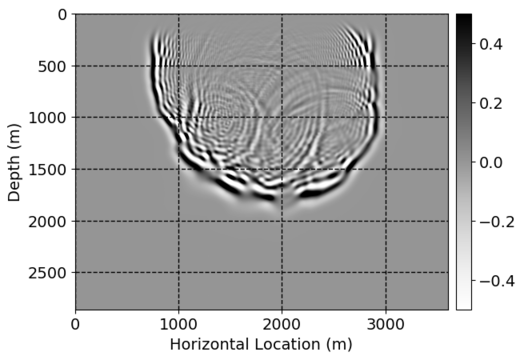}}
\subfloat[\label{compare-10-res-6}]{\includegraphics[width=0.250\hsize]{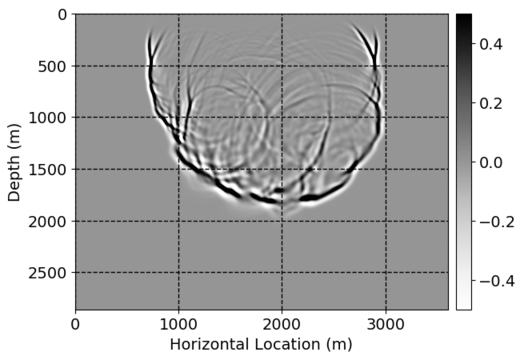}}
\subfloat[\label{compare-10-err-6}]{\includegraphics[width=0.250\hsize]{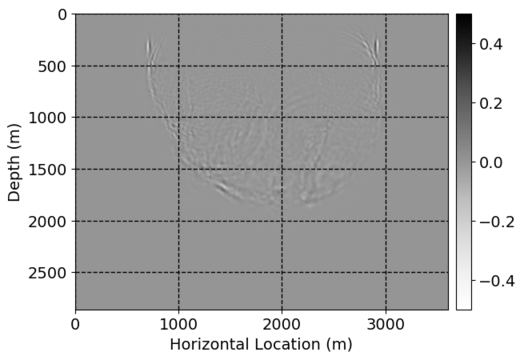}}
\\
\subfloat[\label{compare-10-true-7}]{\includegraphics[width=0.250\hsize]{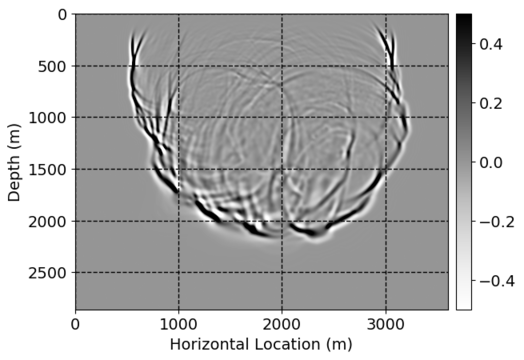}}
\subfloat[\label{compare-10-dsp-7}]{\includegraphics[width=0.250\hsize]{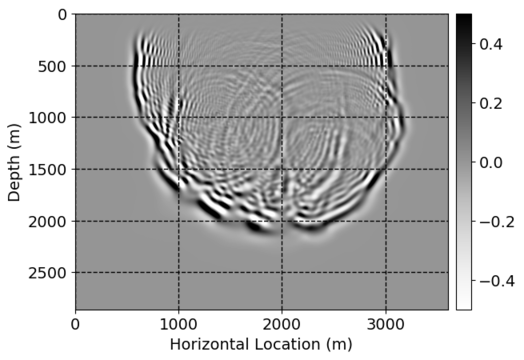}}
\subfloat[\label{compare-10-res-7}]{\includegraphics[width=0.250\hsize]{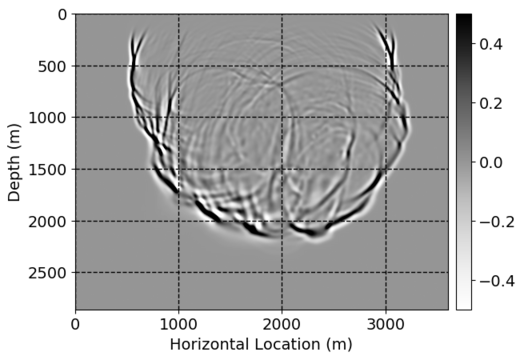}}
\subfloat[\label{compare-10-err-7}]{\includegraphics[width=0.250\hsize]{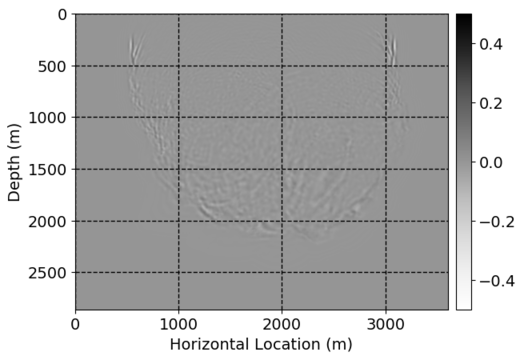}}
\\
\subfloat[\label{compare-10-true-8}]{\includegraphics[width=0.250\hsize]{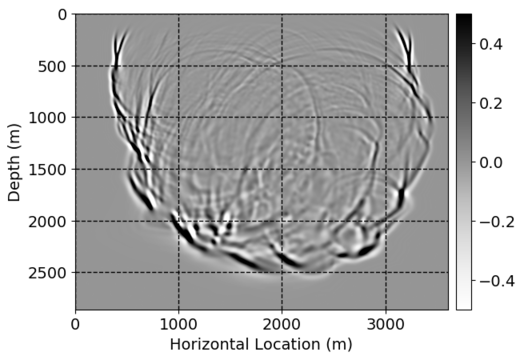}}
\subfloat[\label{compare-10-dsp-8}]{\includegraphics[width=0.250\hsize]{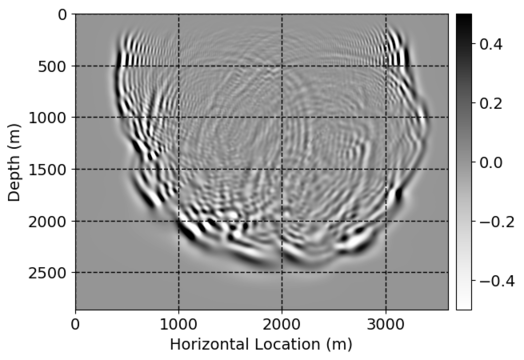}}
\subfloat[\label{compare-10-res-8}]{\includegraphics[width=0.250\hsize]{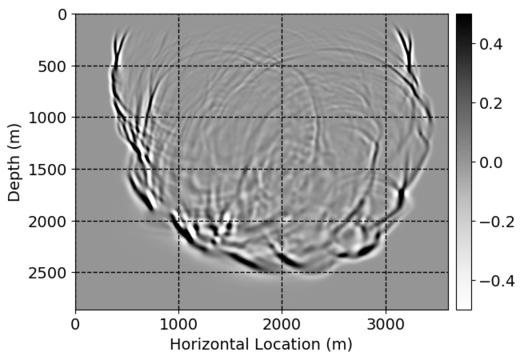}}
\subfloat[\label{compare-10-err-8}]{\includegraphics[width=0.250\hsize]{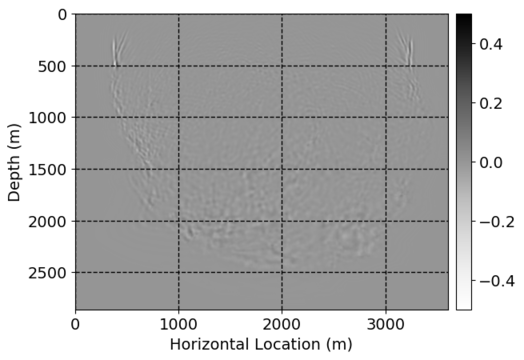}}
\\
\subfloat[\label{compare-10-true-9}]{\includegraphics[width=0.250\hsize]{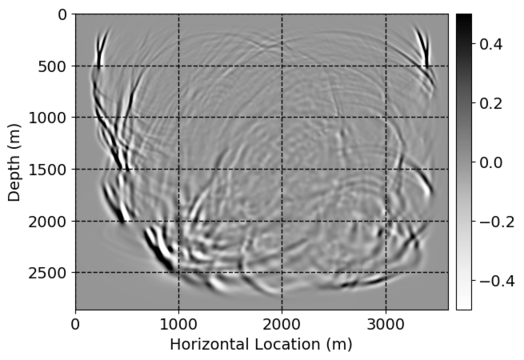}}
\subfloat[\label{compare-10-dsp-9}]{\includegraphics[width=0.250\hsize]{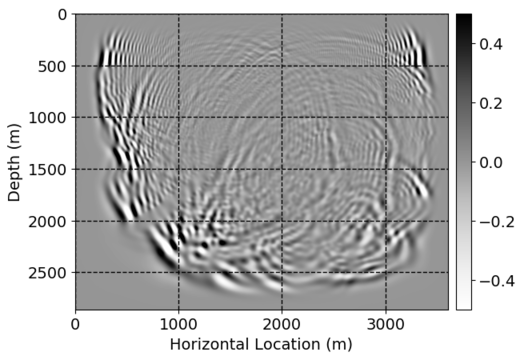}}
\subfloat[\label{compare-10-res-9}]{\includegraphics[width=0.250\hsize]{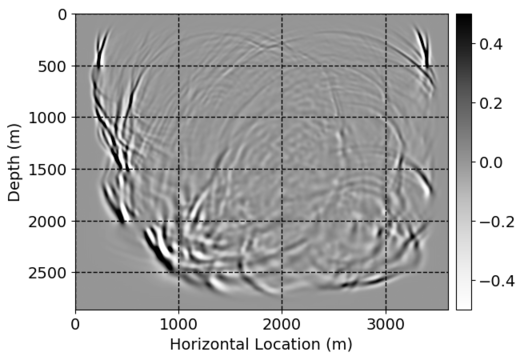}}
\subfloat[\label{compare-10-err-9}]{\includegraphics[width=0.250\hsize]{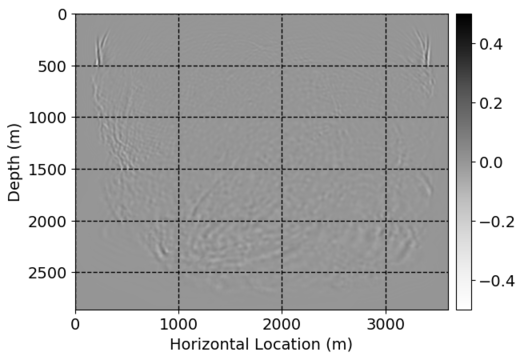}}
\caption{Single CNN low-to-high-fidelity mapping with ten timesteps to
be corrected, second part. First column from top to bottom: a) to q)
high-fidelity wavefield snapshots, in order. Second column from top to
bottom: b) to r) low-fidelity wavefield snapshots simulated by solving
Equation~\ref{low-fidelity} with the same simulation time as
high-fidelity wavefields, in order. Third column from top to bottom: c)
to s) result of single CNN low-to-high-fidelity mapping. Fourth column
from top to bottom: d) to t) difference between first and third column,
in order.}\label{compare-10-2}
\end{figure}

Finally, we will show the wavefield corrected by the single CNN
low-to-high-fidelity mapping method, for comparison with our proposed
method. Figures~\ref{compare-3} and~\ref{compare-5} indicate the
corrected wavefields, for cases where three and five timesteps need to
be corrected. Figures~\ref{compare-10-1} and~\ref{compare-10-2}
demonstrate the corrected wavefields, for the case where ten timesteps
need to be corrected, in two parts. In Figures~\ref{compare-3} $-$
\ref{compare-10-2}, the first, second, third, and fourth columns depict
the high-fidelity and low fidelity wavefield snapshots, corrected
low-fidelity wavefield snapshots, and the error in numerical dispersion
removal, respectively. In each columns, from top to bottom, the
simulation time increases.

For comparison between our proposed method and single CNN
low-to-high-fidelity mapping, compare Figures~\ref{corr-3}
with~\ref{compare-3}, \ref{corr-5} with~\ref{compare-5}, \ref{corr-10-1}
with~\ref{compare-10-1}, \ref{corr-10-2} with~\ref{compare-10-2}. As it
can be seen, the single CNN low-to-high-fidelity mapping method
maintains the quality of its performance when the number of timesteps
that need to be corrected increases. On the other hand, as the number of
CNNs in neural network augmented wave-equation simulation increases, the
performance drops, by keeping the maximum number of iterations fixed.
Also, by comparing Tables~\ref{training-details}
and~\ref{training-details-compare}, we observe that the training time
needed for single CNN low-to-high-fidelity mapping, when number of
timesteps needed to be corrected increases, for fixed number of maximum
iterations, grows very slowly compared to the training time required for
neural network augmented wave-equation simulation.

\section{Conclusions}\label{conclusions}

Our numerical experiments demonstrate that, given suitable training
data, the well-trained neural network augmented wave-equation simulator
is capable of approximating wavefield snapshots simulated by
high-fidelity simulation. In this work, as a proxy of inaccurate
physics, we simulate wave-equation with finite-difference method, using
a poor discretization of Laplacian. Although not computationally
favorable to high-fidelity wave simulation, we showed that the learned
wave simulator deals with inaccurate physics. An important observation
we made is that training time of the proposed method gets quickly very
long and to achieve high accuracy, it may not be possible to utilize too
many CNNs. On the other hand, training time required for the single CNN
low-to-high-fidelity mapping experiments, conducted for the sake of
comparison, grows very slowly as the number of timesteps to be corrected
increases. In future, we intend to initialize the CNN parameters in the
proposed method with the parameters of a CNN trained by the single CNN
low-to-high-fidelity mapping algorithm. The initialization may
significantly reduce the training time needed for the neural network
augmented wave-equation simulation method, and may give the chance to
fine-tune the CNNs to the specific timestep that each CNN is assigned to
correct.

\section{Acknowledgments}\label{acknowledgments}

The authors thank Xiaowei Hu for his open-access repository\footnote{\url{https://github.com/xhujoy/CycleGAN-tensorflow}}
on GitHub. Our software implementation built on this work.

\bibliography{report}

\begin{thebibliography}{15}
\providecommand{\natexlab}[1]{#1}
\providecommand{\url}[1]{\texttt{#1}}
\expandafter\ifx\csname urlstyle\endcsname\relax
  \providecommand{\doi}[1]{doi: #1}\else
  \providecommand{\doi}{doi: \begingroup \urlstyle{rm}\Url}\fi

\bibitem[Ruthotto and Haber(2018)]{ruthotto2018deep}
Lars Ruthotto and Eldad Haber.
\newblock Deep neural networks motivated by partial differential equations.
\newblock \emph{CoRR}, abs/1804.04272, 2018.
\newblock URL \url{http://arxiv.org/abs/1804.04272}.

\bibitem[Raissi(2018)]{raissi2018deep}
Maziar Raissi.
\newblock Deep hidden physics models: Deep learning of nonlinear partial
  differential equations.
\newblock \emph{The Journal of Machine Learning Research}, 19\penalty0
  (1):\penalty0 932--955, 2018.

\bibitem[Moseley et~al.(2018)Moseley, Markham, and
  Nissen-Meyer]{moseley2018fast}
Benjamin Moseley, Andrew Markham, and Tarje Nissen-Meyer.
\newblock Fast approximate simulation of seismic waves with deep learning.
\newblock \emph{arXiv preprint arXiv:1807.06873}, 2018.

\bibitem[Siahkoohi et~al.(2019)Siahkoohi, Louboutin, and
  Herrmann]{siahkoohi2019transfer}
Ali Siahkoohi, Mathias Louboutin, and Felix~J. Herrmann.
\newblock The importance of transfer learning in seismic modeling and imaging.
\newblock 2019.
\newblock Submitted to GEOPHYSICS in February 2019.

\bibitem[Rizzuti et~al.(2019)Rizzuti, Siahkoohi, and
  Herrmann]{rizzuti2019EAGElis}
Gabrio Rizzuti, Ali Siahkoohi, and Felix~J. Herrmann.
\newblock Learned iterative solvers for the helmholtz equation.
\newblock In \emph{EAGE Annual Conference Proceedings}, 06 2019.
\newblock \doi{10.3997/2214-4609.201901542}.
\newblock URL
  \url{https://slim.gatech.edu/Publications/Public/Conferences/EAGE/2019/rizzuti2019EAGElis/rizzuti2019EAGElis.pdf}.
\newblock (EAGE, Copenhagen).

\bibitem[Szegedy et~al.(2017)Szegedy, Ioffe, Vanhoucke, and
  Alemi]{szegedy2017inception}
Christian Szegedy, Sergey Ioffe, Vincent Vanhoucke, and Alexander~A Alemi.
\newblock {I}nception-v4, {I}nception-{R}es{N}et and the {I}mpact of {R}esidual
  {C}onnections on {L}earning.
\newblock In \emph{Proceedings of the Thirty-First Association for the
  Advancement of Artificial Intelligence Conference on Artificial Intelligence
  (AAAI-17)}, volume~4, pages 4278--4284, 2017.
\newblock URL \url{http://aaai.org/ocs/index.php/AAAI/AAAI17/paper/view/14806}.

\bibitem[{Goodfellow} et~al.(2014){Goodfellow}, {Pouget-Abadie}, {Mirza}, {Xu},
  {Warde-Farley}, {Ozair}, {Courville}, and {Bengio}]{Goodfellow2014}
Ian {Goodfellow}, Jean {Pouget-Abadie}, Mehdi {Mirza}, Bing {Xu}, David
  {Warde-Farley}, Sherjil {Ozair}, Aaron {Courville}, and Yoshua {Bengio}.
\newblock {Generative Adversarial Nets}.
\newblock \emph{Advances in neural information processing systems}, pages
  2672--2680, 2014.

\bibitem[Siahkoohi et~al.(2018)Siahkoohi, Louboutin, Kumar, and
  Herrmann]{siahkoohi2018deep}
Ali Siahkoohi, Mathias Louboutin, Rajiv Kumar, and Felix~J. Herrmann.
\newblock Deep-convolutional neural networks in prestack seismic: Two
  exploratory examples.
\newblock \emph{SEG Technical Program Expanded Abstracts 2018}, pages
  2196--2200, 2018.
\newblock \doi{10.1190/segam2018-2998599.1}.
\newblock URL
  \url{https://library.seg.org/doi/abs/10.1190/segam2018-2998599.1}.

\bibitem[Hu et~al.(2019)Hu, Han, Shrivastava, and
  Zwicker]{hu2019render4completion}
Tao Hu, Zhizhong Han, Abhinav Shrivastava, and Matthias Zwicker.
\newblock Render4completion: Synthesizing multi-view depth maps for 3d shape
  completion.
\newblock \emph{arXiv preprint arXiv:1904.08366}, 2019.

\bibitem[Kingma and Ba(2015)]{kingma2015adam}
Diederik Kingma and Jimmy Ba.
\newblock Adam: A method for stochastic optimization.
\newblock In \emph{International Conference on Learning Representations}, 2015.

\bibitem[Johnson et~al.(2016)Johnson, Alahi, and
  Fei-Fei]{johnson2016perceptual}
Justin Johnson, Alexandre Alahi, and Li~Fei-Fei.
\newblock {P}erceptual {L}osses for {R}eal-{T}ime {S}tyle {T}ransfer and
  {S}uper-{R}esolution.
\newblock In \emph{Computer Vision -- European Conference on Computer Vision
  (ECCV) 2016}, pages 694--711. Springer International Publishing, 2016.
\newblock \doi{10.1007/978-3-319-46475-6_43}.
\newblock URL
  \url{https://link.springer.com/chapter/10.1007%2F978-3-319-46475-6_43}.

\bibitem[He et~al.(2016)He, Zhang, Ren, and Sun]{he2016deep}
Kaiming He, Xiangyu Zhang, Shaoqing Ren, and Jian Sun.
\newblock {D}eep {R}esidual {L}earning for {I}mage {R}ecognition.
\newblock In \emph{The IEEE Conference on Computer Vision and Pattern
  Recognition (CVPR)}, pages 770--778, June 2016.
\newblock \doi{10.1109/CVPR.2016.90}.
\newblock URL \url{https://ieeexplore.ieee.org/document/7780459}.

\bibitem[Yosinski et~al.(2014)Yosinski, Clune, Bengio, and
  Lipson]{yosinski2014transferable}
Jason Yosinski, Jeff Clune, Yoshua Bengio, and Hod Lipson.
\newblock How transferable are features in deep neural networks?
\newblock In \emph{Advances in neural information processing systems}, pages
  3320--3328, 2014.

\bibitem[{Louboutin} et~al.(2018){Louboutin}, {Lange}, {Luporini}, {Kukreja},
  {Witte}, {Herrmann}, {Velesko}, and {Gorman}]{devito-api}
M.~{Louboutin}, M.~{Lange}, F.~{Luporini}, N.~{Kukreja}, P.~A. {Witte}, F.~J.
  {Herrmann}, P.~{Velesko}, and G.~J. {Gorman}.
\newblock Devito: an embedded domain-specific language for finite differences
  and geophysical exploration.
\newblock \emph{CoRR}, abs/1808.01995, Aug 2018.
\newblock URL \url{https://arxiv.org/abs/1808.01995}.

\bibitem[{Luporini} et~al.(2018){Luporini}, {Lange}, {Louboutin}, {Kukreja},
  {H{\"u}ckelheim}, {Yount}, {Witte}, {Kelly}, {Gorman}, and
  {Herrmann}]{devito-compiler}
F.~{Luporini}, M.~{Lange}, M.~{Louboutin}, N.~{Kukreja}, J.~{H{\"u}ckelheim},
  C.~{Yount}, P.~{Witte}, P.~H.~J. {Kelly}, G.~J. {Gorman}, and F.~J.
  {Herrmann}.
\newblock Architecture and performance of devito, a system for automated
  stencil computation.
\newblock \emph{CoRR}, abs/1807.03032, jul 2018.
\newblock URL \url{http://arxiv.org/abs/1807.03032}.

\end{thebibliography}

\end{document}